\documentclass[10pt,]{article}
\pdfoutput=1
\usepackage[letterpaper, margin = 1in, footskip = 0.25in]{geometry}
\usepackage{amsmath,amssymb,setspace,mathabx}
\usepackage{graphicx}
\usepackage{lscape}
\usepackage{verbatim}
\usepackage{rotating}
\usepackage{framed}
\usepackage{booktabs}
\usepackage{longtable}
\usepackage{array}
\usepackage{multirow}
\usepackage[table]{xcolor}
\usepackage{wrapfig}
\usepackage{float}
\usepackage{colortbl}
\usepackage{pdflscape}
\usepackage{tabu}
\usepackage{threeparttable}
\usepackage{threeparttablex}
\usepackage[normalem]{ulem}
\usepackage{makecell}
\usepackage{caption}
\usepackage{pdflscape}
\newcommand{\blandscape}{\begin{landscape}}
\newcommand{\elandscape}{\end{landscape}}

\providecommand{\tightlist}{%
  \setlength{\itemsep}{0pt}\setlength{\parskip}{0pt}}

\makeatletter
\def\ScaleIfNeeded{%
  \ifdim\Gin@nat@width>\linewidth
    \linewidth
  \else
    \Gin@nat@width
  \fi
}
\makeatother

\let\Oldincludegraphics\includegraphics
{%
 \catcode`\@=11\relax%
 \gdef\includegraphics{\@ifnextchar[{\Oldincludegraphics}{\Oldincludegraphics[width=\ScaleIfNeeded]}}%
}%

\usepackage{hyperref}

\newcommand{\beginsupplement}{
  \setcounter{table}{0}  
  \renewcommand{\thetable}{S\arabic{table}} 
  \setcounter{figure}{0} 
  \renewcommand{\thefigure}{S\arabic{figure}}
}

\title{Evaluating the role of data quality when sharing information in
hierarchical multi-stock assessment models, with an application to dover
sole}
\author{Samuel D. N. Johnson (corresponding) \and Sean P. Cox}
\date{\today}

\begin{document}


\begin{titlepage}

\begin{flushleft}

\noindent
\textbf{Evaluating the role of data quality when sharing information in
hierarchical multi-stock assessment models, with an application to dover
sole}\\[0.2in]

{Samuel D. N. Johnson (corresponding), Sean P. Cox, }

{School of Resource and Environmental Management, Simon Fraser
University, 8888 University Drive, BC, Canada,\\ }
\vspace{0.2in}

{\href{mailto:samuelj@sfu.ca}{\nolinkurl{samuelj@sfu.ca}},\\ \href{mailto:spcox@sfu.ca}{\nolinkurl{spcox@sfu.ca}},\\ }
\vspace{0.2in}

\textbf{Abstract} \\
An emerging approach to data-limited fisheries stock assessment uses
hierarchical multi-stock assessment models to group stocks together,
sharing information from data-rich to data-poor stocks. In this paper,
we simulate data-rich and data-poor fishery and survey data scenarios
for a complex of dover sole stocks. Simulated data for individual stocks
were used to compare estimation performance for single-stock and
hierarchical multi-stock versions of a Schaefer production model. The
single-stock and best performing multi-stock models were then used in
stock assessments for the real dover sole data. Multi-stock models often
had lower estimation errors than single-stock models when assessment
data had low statistical power. Relative errors for productivity and
relative biomass parameters were lower for multi-stock assessment model
configurations. In addition, multi-stock models that estimated
hierarchical priors for survey catchability performed the best under
data-poor scenarios. We conclude that hierarchical multi-stock
assessment models are useful for data-limited stocks and could provide a
more flexible alternative to data-pooling and catch only methods;
however, these models are subject to non-linear side-effects of
parameter shrinkage. Therefore, we recommend testing hierarchical
multi-stock models in closed-loop simulations before application to real
fishery management systems. \\[0.2in]
\end{flushleft}

\today

\end{titlepage}

\setcounter{page}{2}


\newpage

\hypertarget{introduction}{%
\section{Introduction}\label{introduction}}

Fisheries stock assessment modeling uses catch and abundance monitoring
data to estimate the status and productivity of exploited fish stocks
(Hilborn 1979). Despite improvements in catch monitoring and increasing
prevalence and quality of fishery-independent surveys of abundance, many
fisheries remain difficult to assess because the data lack sufficient
statistical power to estimate key quantities necessary for management
(Peterman 1990). Low power data may arise, for example, because
time-series are short relative to the productivity cycles of exploited
fish stocks, historical fishing patterns may be weak or uninformative,
and monitoring data may simply be too noisy to extract biomass and
productivity signals (Magnusson and Hilborn 2007). Where these
situations occur, stocks are often deemed \emph{data-limited} (MacCall
2009, Carruthers et al. 2014).

An emerging approach to fisheries stock assessment is to use a
hierarchical approach to assess data-limited stocks simultaneously with
data-rich stocks. Data-limited stocks can ``borrow information'' from
data-rich stocks, providing a compromise between data-intensive
single-stock assessments and problematic data-pooling approaches (Jiao
et al. 2009, 2011, Punt et al. 2011). The hierarchical multi-stock
approach, which shares information between data-rich and data-poor
stocks, treats multiple stocks of the same species as replicates that,
to varying degrees, share environments, life history characteristics,
ecological processes, and fishery interactions (Peterman et al. 1998,
Punt et al. 2002, Malick et al. 2015). Information present in the
observations for data-rich replicates is shared with more data-poor
replicates via hierarchical prior distributions on parameters of
interest (Punt et al. 2011, Thorson et al. 2015). Sharing information in
this way could improve scientific defensibility of assessments for
data-limited stocks, because stock status and productivity estimates are
informed by data rather than strong \emph{a priori} assumptions on
population dynamics parameters.

Information-sharing properties of hierarchical models are realized as
the shared hierarchical priors induce shrinkage of estimated parameters
towards the overall prior mean (Carlin and Louis 1997, Gelman et al.
2014). Although shrinkage can reduce bias in the presence of high
uncertainty (e.g.~very data-limited stocks), it may also increase bias
for data-rich replicates by pulling estimated parameters closer to the
group mean. Shrinkage properties are well understood for hierarchical
linear models (James and Stein 1961, Raudenbush and Bryk 2002),
including those applied in fisheries. For example, when estimating
productivity of Pacific salmon stocks, hierarchical Ricker
stock-recruitment models are more successful at explaining variation in
stock productivity when stocks are grouped at scales consistent with
climatic variation (Peterman et al. 1998, Mueter et al. 2002). It is
unclear, however, whether the benefits observed for linear models extend
to iteroparous groundfish stocks, for which productivity parameters are
deeply embedded within non-linear population dynamics and statistical
models.

Parameter shrinkage has been observed in stock assessments for
data-limited groundfish and shark species when grouped with
data-moderate species (Jiao et al. 2009, 2011, Punt et al. 2011), but it
is unknown whether such shrinkage in reality increases or decreases bias
in parameter estimates. Simulation tests of the hierarchical multi-stock
approach to age-structured assessments revealed that bias reductions in
one species often induce greater bias for others in the assessment
group, indicating that shrinkage could imply unwanted trade-offs (Punt
et al. 2005).

In this paper, we used a simulation approach to investigate
relationships between hierarchical model structure, bias, and precision
for hierarchical multi-stock Schaefer stock assessment models. For the
hierarchical multi-stock models, we defined shared prior distributions
on survey catchability and optimal harvest rate (productivity) and then
identified combinations of shared priors that produced the most reliable
estimates of key management parameters when fit to simulated data from
high and low data quality multi-stock complexes. Best performing singl
and multi-stock models were then applied to real data for a dover sole
complex in British Columbia, Canada.

\hypertarget{methods}{%
\section{Methods}\label{methods}}

We simulated a multi-stock complex representing the dover sole
(\emph{Microstomus Pacificus}) fishery in British Columbia, Canada.
Dover sole stocks were simulated under low to high data quality
(statistical power) scenarios. Under each scenario, bias and precision
metrics were determined for key management parameters under both
single-stock and hierarchical multi-stock Schaefer models. In our
hierarchical multi-stock assessment models shared evolutionary history
and a common scientific survey influenced our choice of shared prior
distributions. For example, stocks that share evolutionary history may
have similar productivity at low stock sizes (Jiao et al. 2009, 2011),
and a common trawl survey may induce correlations in catchability (trawl
efficiency) observation errors.

\hypertarget{study-system}{%
\subsection{Study system}\label{study-system}}

British Columbia's dover sole complex is divided into three distinct but
connected stocks (Figure 1), distributed along the BC coast from the
northern tip of Haida Gwaii, south through Hecate Strait into Queen
Charlotte Sound, and on the west coast of Vancouver Island. Although the
dover sole fishery has operated since 1954, prior to 1970 it was very
limited, increasing to present levels by the late 1980's (Figure 2).

Despite a long history of exploitation, dover sole stocks have never
been evaluated using model-based assessments. No observational data
exists for the Queen Charlotte Sound (QCS) and west coast of Vancouver
Island (WCVI) stocks prior to 2003, precluding a model based assessment
before that time (Fargo 1999). The Haida Gwaii and Hecate Strait (HS)
stock was surveyed from 1984 - 2003 (Figure 2, Survey 1), but data was
only used to perform catch curve analyses for total mortality rate
estimates (Fargo 1998). During 1984 - 2003, a fine-mesh trawl survey was
used for the Vancouver Island stock and a portion of the Hecate Strait
stock, but the survey was not designed for groundfish and produced stock
indices that were highly variable. Since 2003, a new bottom trawl survey
has operated coast-wide, which samples all three stocks (Figure 2,
Survey 2), but no assessment has been performed in that time.

Dover sole may be suitable for a hierarchical multi-stock assessment for
3 main reasons. First, the Hecate Strait stock has longer series of
informative data than the other stocks, potentially providing
information for the other two stocks. Second, modeling a single-species
makes it likely that stock productivities and responses to the
environment are similar. Lastly, all stocks are observed by Survey 2,
making it likely that the observation model parameters for each stock
are similar for that survey. By applying the hierarchical multi-stock
approach, the similarities between stocks may be exploited to the
benefit of the whole complex, extending model based stock assessments
for dover sole for the first time.

\hypertarget{simulation-framework}{%
\subsection{Simulation Framework}\label{simulation-framework}}

Our simulation framework was composed of an operating model that
simulated biological dynamics, catch, and observational data, and an
assessment model that performed both single-stock and hierarchical
multi-stock assessments from the simulated data. Both operating and
assessment models used a process-error Schaefer formulation for biomass
dynamics, where the biomass in each year is deviated from the expected
value using a log-normal process error term. This choice allowed us to
focus on the effects of hierarchical estimation and shrinkage without
confounding among hierarchical priors and the model structure. We used
the R statistical software package to specify the operating model, and
the Template Model Builder (TMB) package to specify the assessment model
(Kristensen et al. 2015, R Core Team 2015).

The simulation approach is described below in 3 main sections (i) the
operating model, (ii) assessment models, and (iii) simulation
experiments. The next section describes the operating model structure,
including process errors, and how catch and survey observations were
generated. Assessment models are then outlined, with details of the
shared hierarchical prior distributions given in the supplemental
material. Finally, we present the experimental design and performance
metrics for the simulations.

\hypertarget{operating-model}{%
\subsubsection{Operating model}\label{operating-model}}

We simulated biomass dynamics for each stock \(s\) in our assessment
complex on an annual time step \(t\), using the process-error Schaefer
model (Punt 2003) \begin{equation}
B_{s,t+1} = \left(B_{s,t} + r_sB_{s,t} \left( 1 - B_{s,t}/B_{s,0} \right) - C_{s,t} \right) e^{\epsilon_{s,t}}{,}
{\label{eq:OM}}
\end{equation} \noindent where \(B_{s,t}\) is the biomass of stock \(s\)
at time \(t\), \(r_s\) is the intrinsic rate of increase, \(B_{s,0}\) is
the unfished equilibrium biomass, and \(\epsilon_{s,t}\) is the process
error deviation for stock \(s\) at time \(t\). Schaefer model process
error deviations \(\epsilon_{s,t}\) were decomposed via the sum of a
shared (across stocks) mean year-effect \(\bar\epsilon_t\), and a
correlated (among stocks) stock-specific effect \(\zeta_{s,t}\), which
is the \(s\) component of the vector \(\zeta_{\cdot,t}\), that is, \[
\begin{split} 
\epsilon_{s,t}  & =     \bar{\epsilon}_t + \zeta_{s,t}{,} \\
\bar\epsilon_t  &\sim   N(0,\kappa){,} \\
\zeta_{\cdot,t} &\sim   N(\vec{0},\Sigma){.}
\end{split}
\] \noindent We specified the covariance matrix \(\Sigma\) as the
diagonal decomposition \(\Sigma = DMD\), where \(D\) is a diagonal
matrix of stock-specific standard deviations \(\sigma_s\), and \(M\) is
the matrix of stock correlations. For simplicity, we simulated all
stocks with identical pair-wise covariances, i.e., for a 3 stock complex
\[
M = \left( \begin{array}{ccc}
              1 & 0.5 & 0.5 \\
              0.5 & 1 & 0.5 \\
              0.5 & 0.5 & 1
              \end{array}  \right),
\] and all stocks experienced the same magnitude of stock-specific
process errors where \(\sigma_s = \sigma\), implying \[
D = \left( \begin{array}{ccc}
              \sigma & 0 & 0 \\
              0 & \sigma & 0 \\
              0 & 0 & \sigma
              \end{array}  \right).
\] The operating model values of \(\kappa\) and \(\sigma\) were chosen
to give a total process error variance of
\(\sigma^2 + \kappa^2 = 0.01\), or roughly a 10\% total relative
standard error (Table 1).

We simulated 34 years of fishery history from 1984 (\(t = 1\)) to 2017
(\(t = 34\)). Each stock was initalized in 1984 at a pre-determined
depletion level \(d_{s,1}\) relative to unfished biomass, i.e.,
\(B_{s,1} = d_{s,1} \cdot B_{s,0}\). Unless otherwise stated, we set
\(d_{s,1} = 1\), which is varied as an experimental factor (Table 2).
Because we simulated a single-species, multi-stock complex, we used the
same base biological parameters \(B_{s,0}\) kilo-tonnes, and \(r_s\) for
all stocks \(s\) (Table 1). While identical parameters may not
adequately represent the true dover sole complex, it helped us focus on
the effects of shrinkage in parameter estimates, rather than differences
in biological parameters. This choice also simplified reporting and
interpretation of the results, allowing us to focus on parameter
estimates for a smaller set of representative stocks, rather than
analysing every stock in the complex.

Fishery catch and fishery independent biomass indices were sampled from
each stock each year. We simulated perfectly implemented catch
\(C_{s,t} = U_{s,t}B_{s,t}\), where \(U_{s,t}\) was the harvest rate
applied in a pulse fishing event following each year's production. We
also assumed that catch was fully observed (i.e., no under-reporting).
Harvest rates were simulated in three temporal phases and scaled to
optimal fishing mortality as \(U_{s,t} = U^{mult}_t \cdot U_{s,MSY}\),
where \(U^{mult}_t\) is the piecewise linear function of \(t\):
\begin{equation}
U^{mult}_t = \left\{ 
\begin{array}{ll} 
U_i + (t-1) \cdot \frac{U_{d} - 0.2}{t_{d} - 1}& 1 \leq t \leq t_d{,} \\
U_{d} + (t - t_{d}) \cdot \frac{U_{m} - U_{d}}{t_{m} - t_{d}} & t_d \leq t \leq t_m{,} \\
U_{m} & t_{m} \leq t \leq T{;}
\end{array} \right.
\label{eq:Fhist}
\end{equation} \noindent where \(U_i\), \(U_{d}\) and \(U_{m}\) are the
initial, development, and managed phase harvest rates, respectively,
\(t_{d}\) is the last time step of the development phase, and \(t_{m}\)
is the beginning of the final managed phase (Figure 3). In the base
operating model, we used values \(U_i = 0.2\), \(U_d = 4\) and
\(U_m = 1\) for harvest rate multipliers, with \(t_d = 5\) (1988), and
\(t_m = 15\) (1998) for phase timing, to simulate a high initial
development phase followed by a reduction in pressure, allowing the
stock to recover. This formulation was designed to create more and less
informative catch histories, depending on the parameter values (Schnute
and Richards 1995).

Survey indices of biomass were simulated for each stock \(s\) and survey
\(o\) via the observation model \[
I_{o,s,t} = q_{o,s}B_{s,t}e^{\delta_{o,s,t}},
\] \noindent where \(q_{o,s}\) is stock-specific catchability
coefficient for survey \(o\). Observation errors were simulated via the
distribution \[
\delta_{o,s,t} \sim N(0,\tau_o),
\] \noindent where \(\tau_o\) is the survey observation error log scale
standard deviation for survey \(o\). Within each survey, stock-specific
catchabilities \(q_{o,s}\) were randomly drawn from a log-normal
distribution with a mean survey catchability coefficient \(\bar{q}_o\)
and between-stock log-standard deviation \(\iota_{q,o}\) via \[
q_{o,s} \sim \log N( \bar{q}_o, \iota_{q,o}).
\]

It is not always the case that catchability will be correlated closely
between stocks. Indeed, we were able to model catchability as a
correlated process between stocks because we used swept area biomass
estimates as our stock indices. To see this, note that the general
formula for catchability is \(q = c a/A\), where \(c\) is gear
efficency, \(a\) is the average area fished by the gear during the
survey, and \(A\) is the total area of the surveyed stock's habitat
(Arreguı'n-Sánchez 1996). Because the geographic boundaries of stocks
may differ, it will usually be the case that \(A \neq A'\) between 2
distinct stocks \(s\) and \(s'\), even if the average surveyed area
\(a\) and gear efficiency \(c\) are the same. For a trawl survey, it is
advantageous that the area swept by the fishing gear is often known
exactly, with \(a = t \cdot v \cdot w\), where \(t\) is the standard tow
duration, \(v\) is the tow velocity and \(w\) is the door-width of the
trawl net. Therefore, the total of randomly sampled survey catches
\(C_t = qE_tB_t\) from a total effort of \(E_t = n_t\) tows can be
transformed into biomass estimates when scaled by the reciprocal of the
proportion of area swept, e.g. \(B'_t = \frac{A}{n_t a}C_t = cB_t\).
Then the effect of stock area is scaled out of the index, and
catchability is reduced to gear efficency \(c\), or the response of
individual fish to the survey gear. We then assumed that this response
is similar between individuals of the same species. This calculation
extends to swept area biomass estimates calculated from a stratified
survey, like the trawl survey used for Dover Sole.

We simulated biomass indices from two surveys operating over different
periods to emulate the current dover sole complex history (Figure 2).
The first (\(o = 1\)) represented Survey 1, which operated from 1984 to
2003 (\(t = 1,\dots,20\)), with observation model parameters
\(\tau_1 = 0.2\) for the observation errors, and a mean survey
catchability of \(\bar{q}_1 = 0.5\) with a standard deviation of
\(\iota_{q,1} = 0.1\). For survey 2 (\(o = 2\)), which operated from
2003 to 2017 (\(t = 20,\dots,34\)), we modeled an observation error
standard deviation of \(\tau_2 = 0.4\), and a mean catchability of
\(\bar{q}_2 = 0.6\) with a standard deviation of \(\iota_{q,2} = 0.1\).

\hypertarget{assessment-model}{%
\subsubsection{Assessment model}\label{assessment-model}}

We estimated stock-specific biological and management parameters using
multi-stock and single-stock versions of a state-space Schaefer stock
assessment model. We minimized the effect of assessment model
mis-specification by matching the deterministic components of the
biomass dynamics in the assessment models and the operating model,
Equation \eqref{eq:OM}. Details of the assessment model prior
distributions are not presented in this section. Instead, the equations
for each multi-level prior in the hierarchical multi-stock assessment
model are given in Table 3, and the details of all prior distributions
are given in supplementary material S1.

\hypertarget{hierarchical-multi-stock-assessment-models}{%
\paragraph{Hierarchical multi-stock assessment
models}\label{hierarchical-multi-stock-assessment-models}}

For the full hierarchical multi-stock model, we defined shared prior
distributions on (1) conditional maximum likelihood estimates of
stock-specific catchability \(\hat{q}_{o,s}\) within each survey and (2)
optimal harvest rate \(U_{s,MSY}\), which was used as a surrogate for
stock productivity (Table 3). In total, we defined 4 configurations,
including a ``null'' multi-stock model. Each multi-stock model
configuration was defined by whether each of the hierarchical priors was
estimated along with the leading model parameters. When a hierarchical
prior was ``off'', shared priors were bypassed and the model used the
fixed hyperprior mean and standard deviation instead (Table 3, Single
level priors). Full details of the single and multi-level priors are in
supplemental material.

\hypertarget{single-stock-assessment-model}{%
\paragraph{Single-stock assessment
model}\label{single-stock-assessment-model}}

The single-stock assessment model was defined as a special case of the
multi-stock null model. Prior distributions on catchability and
productivity were the single level priors (Table 3, q.4 and U.4).

\hypertarget{optimization}{%
\paragraph{Optimization}\label{optimization}}

Assessment models applied the Laplace approximation to integrate the
objective function over random effects, obtaining a marginalized
likelihood (Kristensen et al. 2015). The marginalized likelihood was
then maximized via the \texttt{nlminb()} function in \texttt{R} to
produce parameter estimates and corresponding asymptotic standard errors
(R Core Team 2015). We considered an assessment model converged when the
optimisation algorithm reported convergence, which was characterized by
gradient components of the TMB model all having magnitude less than
0.0001, and a positive definite Hessian matrix. Standard errors of
derived parameters were estimated from the Hessian matrix using the
delta method. The estimated process errors \(\zeta_{s,t}\) were treated
as random effects for all model configurations, and stock-specific
catchability parameters \(\log q_{os}\) were treated as random effects
when the shared catchability prior was estimated.

\hypertarget{simulation-experiments}{%
\subsection{Simulation experiments}\label{simulation-experiments}}

We used an experimental design approach to investigate performance of
the four hierarchical multi-stock assessment model configurations under
different levels of statistical power in the simulated data. Multiple
scenarios were used to determine whether (and possibly to what extent)
hierarchical multi-stock assessment methods could provide better
estimates of key management parameters, compared to single-stock
approaches, when fitted to data with low statistical power.

Experimental factors were selected to increase and decrease the
statistical power, or quality, of the simulated assessment data. The
choice of factors determining high- and low-information scenarios was
guided by previous studies of assessment models, as well as our own
experience with production model behaviour (Hilborn 1979, Magnusson and
Hilborn 2007, Cox et al. 2011). Combinations of experimental factors
were chosen according to a space-filling experimental design (Table S1)
(Kleijnen 2008). Space filling designs improve the efficiency of large
simulation experiments by reducing the number of individual runs, while
still producing acceptable estimates of factor effects.

We represented high and low statistical power scenarios by varying 5
experimental factors: (1) historical fishing intensity; (2) the number
\(S\) of stocks in the complex; (3) the number \(L\) of low information
stocks in the complex; (4) the initial year of stock assessment \(T_1\)
for the \(L\) low information stocks; and (5) the initial stock
depletion levels \(d_{s,1}\) for the \(L\) low information stocks (Table
2).

We defined 2 levels of historical fishing intensity, which modified
\(U_i\), \(U_d\) and \(U_m\) in Equation \eqref{eq:Fhist}. Levels were
chosen to produce one-way and two-way trip dynamics when the simulated
biomass was initialized at unfished equilibrium in 1984. One way trips
were produced by fishing at a constant rate of \(U_{s,MSY}\) for the
whole historical period (top row, Figure 3), while the two-way trips
were produced by the base operating model settings (bottom row, Figure
3). The constant harvest rate scenarios had two significant
disadvantages: first, it is impossible, in general, to estimate the
optimal harvest rate without overfishing (Hilborn and Walters 1992, Ch
1), which does not occur in these scenarios; second, when stocks were
initialized at fished levels it was difficult to determine the stock
size and initial biomass.

Complex sizes \(S\) were chosen to test the intuitive notion that
grouping more stocks together increases the benefit of shrinkage. We
tested the sensitivity of this notion to relative differences in the
number of stocks via the factor \(L\), which determined how many of the
\(S\) stocks were ``low information''. Low information stocks had short
time series and fished initialisation at a pre-determined relative
biomass level, which together reduced or removed the contrast in the
biomass dynamics and lower the quality of observational data. By
initializing the assessments of low information stocks when Survey 2 was
initiated, and simulating Survey 2 as a shorter and noisier series of
observations, we subjected those stocks to non-equilibrium starting
conditions as well as poor quality survey data, a situation that is
likely common for data-limited fisheries. When \(L > 0\), we estimated
the initial biomass \(B_{s,T_1}\) for the low information stocks in
addition to unfished biomass, optimal harvest rate and catchability.

We fit the single-stock and each hierarchical multi-stock assessment
model configurations to simulated data under each combination of
experimental factors. The distributions used for the single-level and
multi-level hyperpriors (Table 3, q.2, q.4, U.2, and U.4) were given
random mean values \(m_q\) and \(m_U\) in each simulation replicate,
chosen from a log-normal distribution centred at the true mean value
(across stocks, and possibly surveys) with a 25\% coefficient of
variation. This randomisation was used to test the robustness of the
assessment model to uncertainty in the prior distribution. The same
initial seed value \(R\) was used across all experimental treatments so
that variability in assessment error distributions was predominantly
affected by the factor levels and model configurations, rather than
random variation in the process and observation errors. Random variation
was not completely avoidable, though, as assessment models would fail to
converge for some combinations of treatment and random seed values. In
these cases we restarted the optimisation with jittered initial
parameter values up to 20 times, after which we moved on to a different
random seed value. The total number of replicates for each experiment
and prior configuration are shown in Table S1.

\hypertarget{performance-metrics}{%
\subsubsection{Performance metrics}\label{performance-metrics}}

We measured performance of both the single-stock and multi-stock
assessment models by their ability to estimate current biomass
\(\hat{B}_{s,2017}\), MSY level biomass \(\hat{B}_{s,MSY}\), equilibrium
optimal harvest rate \(\hat{U}_{s,MSY}\), and relative terminal biomass
\(\hat{B}_{s,2017}/\hat{B}_{s,0}\). We also found catchability estimates
\(\hat{q}_{o,s}\) to be important in the analysis of these models, so we
calculated performance metrics for catchability as well.

It is important to understand the effect of shrinkage on the bias and
precision of estimates of the key parameters \(\theta\) above, because
such shrinkage may result in misleading harvest advice. For example,
shrinkage may simultaneously increase both bias and precision for a
given parameter (e.g. \(MSY\)), leading to confidence intervals that may
not contain the true parameter value. Therefore, we used four
performance metrics to represent these effects: (1) median relative
errors (MREs); (2) ratios of median absolute relative errors (MAREs);
(3) confidence interval coverage probability (IC); and (4) the
predictive quantile. All metrics are defined in detail below. While MREs
only indicate model bias, all other metrics are affected by both the
bias and precision of the estimator, and can be better interpreted when
the bias is known.

For MRE and MARE metrics, we calculated relative errors
\(RE(\hat\theta_{i,s})\) of the model estimate \(\hat\theta_{i,s}\) for
each replicate \(i\) and stock \(s\), i.e. \[ 
RE (\hat{\theta}_{i,s}) = 100 \cdot \left(\frac{\theta_{i,s} - \hat{\theta}_{i,s}}{\theta_{i,s}}\right).
\] Estimator bias and precision were quantified by computing the median
relative error
\(MRE(\theta_{s}) = \mbox{med}(RE(\hat\theta_{\cdot,s}))\) and median
absolute relative error
\(MARE(\theta_{s}) = \mbox{med}(|RE(\hat\theta_{\cdot,s})|)\) of
relative error distributions \(RE(\hat\theta_{\cdot,s})\) over all
replicates \(i\). We chose to use MAREs because they are independent of
scale and less sensitive to outliers than root mean square errors.
Values closer to zero indicate better performance for both metrics, with
lower MRE values indicating lower bias, and lower MARE values indicating
lower bias, higher precision, or both.

In the simulation experiments we compared assesment models via ratios of
single-stock to multi-stock MARE statistics for each stock \(s\) and
parameter \(\theta\), i.e., \begin{equation}
\Delta(\theta_s) = \frac{MARE_{ss}(\theta_{s})}{MARE_{ms}(\theta_{s})} - 1,
\end{equation} \noindent where \(ss\) and \(ms\) represent the MARE
values for the single- and multi-stock hierarchical assessment model
estimates, respectively. Using this definition, \(\Delta(\theta_s) > 0\)
occured when the multi-stock assessment model had a lower MARE value,
indicating that multi-stock estimates had higher precision, lower bias,
or both. Estimation performance for an assessment complex as a whole was
indicated by an aggregate MARE ratio \(\overline{\Delta}(\theta_s)\) for
each stock's parameter \(\theta_s\), i.e., \[ 
\overline{\Delta}(\theta) = \frac{\sum_s MARE_{ss}(\theta_s)}{\sum_s MARE_{ms}(\theta_s)} - 1,
\] which allowed us to compare estimation performance of single and
multi-stock assessment models over the whole assessment complex.

Interval coverage probability was calculated across reps \(i\) within
each combination of experimental factors and model configuration. We
calculated the realized interval coverage probability under an
assumption of normality on the log scale, because all quantities of
interest are constrained to be positive, and chose the nominal coverage
probability as 50\%, with a corresponding \(z\)-score of 0.67. These two
choices defined our interval coverage probability metric as \[
IC_{50}(\log\theta_s) = \frac{1}{100} \sum_i I(\log\theta \in (\hat{
\log\theta}_{i,s} - 0.67\hat{se}(\log\theta)_{i,s}, \hat{\log\theta}_{i,s} + 0.67\hat{se}(\log\theta)_{i,s})),
\] where \(I\) is the indicator function, \(\hat{\log\theta}_i\) is the
model estimate of \(\log\theta\) in replicate \(i\), and
\(\hat{se}(\log\theta)_i\) is the model standard error of \(\log\theta\)
in replicate \(i\). For a 50\% interval coverage, realized rates
\(IC_{50\%}(\log\theta_s)\) closer to the nominal rate 0.5 are better.
The confidence interval is considered conservative when realized
coverage rates are above the nominal rate, which could indicate either
decreased bias of the parameter estimate or high uncertainty (larger
standard errors). On the other hand, the confidence interval is
considered permissive when realized rates are below the nominal rate,
indicating that the uncertainty may be under-represented by the
parameter estimate and its standard error.

Finally, for each parameter we calculated the distribution of predictive
quantiles over replicates \(i\), defined as \[
Q(\log\theta_{i,s}) = P(\hat{\log\theta_{i,s}} < \log\theta_{i,s}) = \int_{x = -\infty}^{x = \log\theta_{i,s}} f(x ~|~ \hat{\log\theta_{i,s}},\hat{se}(\log\theta)_{i,s})dx,
\] where \(f(x|m,s)\) is the normal probability density function with
mean \(m\) and standard deviation \(s\). The resulting distribution of
quantiles is best interpreted graphically, and indicates how well the
model is estimating parameter uncertainty. Well performing estimators
will have a near-uniform distribution of \(Q\) values, because true
values should be distributed randomly across the full domain of the
parameter's sampling distribution. Estimators that under-represent
uncertainty by produce standard errors that are too small and will,
therefore, have excess density near \(Q = 0\) and \(Q = 1\) (i.e a
\(\bigcup\)-shaped graphical distribution), indicating that true values
have larger \(z\)-scores in the sampling distribution. Models that
over-represent uncertainty have standard errors that are too large and
will collect density near \(Q = .5\) (i.e.~a \(\bigcap\)-shaped
graphical distribution), indiciating lower \(z\)-scores of true values
in the sampling distribution.

We used an experimental design approach for simulation models to analyse
the effects of experimental factors and assessment model configurations
on the MARE and \(\Delta\) performance metrics (Kleijnen 2008). This
method attmpts to simplify the complex response surfaces via a
generalized linear meta-model of teh response surface to simulation
model inputs (i.e.~factor levels and assessment model prior
configurations)(McCullagh 1984). Meta-models are defined in the
supplemental material.

\hypertarget{assessment-for-british-columbia-dover-sole}{%
\subsection{Assessment for British Columbia dover
sole}\label{assessment-for-british-columbia-dover-sole}}

We fit all 8 multi-stock assessment model configurations and the
single-stock assessment model to the dover sole data for the three
stocks in Figure 2. We initialized all stocks in a fished state,
beginning in 1984 for the HS stock, and 2003 for both QCS and WCVI
stocks.

For the prior on \(B_{s,MSY}\) and \(B_{s,init}\), we used a prior mean
value of \(m_{B,s} = 20\) and \(s_{B,s} = 20\), keeping the relative
standard deviation at 100\%. For the process error variances, we tested
two hypotheses for the strength of environmental effects on population
dynamics. These were implemented as choices for the \(\beta\) parameters
of the inverse-gamma prior distributions on process error variance
terms, when using \(\alpha_\sigma = 3\). The first choice was to use
\(\beta_\sigma = 0.16\), placing the prior mode at around \(0.04\),
favouring process errors with a larger standard deviation around
\(\sigma = 0.2\). The second was to use \(\beta_\sigma = 0.01\),
reducing the prior mode to \(0.0025\), favouring process errors with a
small standard deviation around \(\sigma = 0.05\).

For each model fit, we calculated Akaike's information criterion, which
we corrected for the sample size (number of years of survey data) for
each stock (AICc) (Burnham and Anderson 2003). We then selected the
group of multi-stock configurations that performed the best under both
hypotheses according to their AICc values, and present estimates of
optimal harvest rate \(U_{s,MSY}\), terminal biomass \(B_{s,T}\),
optimal biomass \(B_{s,MSY}\), relative biomass \(B_{s,T}/B_{s,0}\), and
current fishing mortality relative to the optimal harvest rate
\(U_{s,T}/U_{s,MSY}\), as well as standard errors for all estimates. We
used the sum of single-stock AICc values to represent the complex
aggregate AICc score for comparing single-stock and multi-stock model
fits. While this may be a slight deviation in use of the AIC, we believe
it is both useful and satisfies the restrictions of the AICc, i.e., the
collection of single-stock models is fit to the same data as the
multi-stock models, and the process of adding AICc values is analogous
to adding single-stock model log-likelihood values within a joint
likelihood.

\hypertarget{results}{%
\section{Results}\label{results}}

When discussing experimental results, we restrict our attention to stock
\(s = 1\), a low information stock if \(L>0\) in the information
scenarios, and identical to the remaining stocks otherwise. We intially
focus on the meta-model effects on MARE ratios \(\Delta(\theta_s)\) and
complex aggregate \(\overline\Delta(\theta)\) to interpret model
configuration effects, and use the remaining metrics to help interpret
factor effects.

\hypertarget{single-stock-versus-multi-stock-assessments-of-the-base-operating-model}{%
\subsection{Single-stock versus multi-stock assessments of the base
operating
model}\label{single-stock-versus-multi-stock-assessments-of-the-base-operating-model}}

As expected, shrinkage effects from hierarchical multi-stock assessment
models often improved precision of key management parameter relative
errors from multi-stock models compared to single-stock models, when fit
to data from the base operating model (Figure 4). Although this pattern
extended across most model configurations and variables, the effect was
most noticeable for optimal harvest rate \(U_{MSY}\) and optimal biomass
\(B_{MSY}\), and weakest for absolute \(B_T\) and relative \(B_T/B_0\)
terminal biomass. Also, the effects of hierarchical priors were most
noticeable for parameters that were subject to those priors,
i.e.~catchability had larger increases in precision under a model
configurations that estimated a shared prior on catchability (Figure 4,
\(q_1,q_2\) under the \(q\) AM configuration).

We found that estimator bias was less sensitive to hierarchical
multi-stock configurations, with sometimes very subtle effects. For
example, for optimal harvest rate \(U_{MSY}\), optimal biomass
\(B_{MSY}\), and survey 1 catchability \(q_1\) estimates were all
relatively unbiased under the single-stock model, and all multi-stock
model configurations had a negligible effect on the bias (Figure 4). In
contrast, survey 2 catchability \(q_2\), and absolute and relative
terminal biomass \(B_T\) and \(B_T/B_0\) were biased under the
single-stock model, so were themselves very sensitive. As with
precision, the bias of catchability \(q_2\) was most reduced by the
\(q\) and \(q/U_{MSY}\) configurations, and these improvements
translated directly into reductions in absolute bias of the terminal
biomass estimates \(B_T\) and \(B_T/B_0\).

The other performance metrics indicated that the \(q\) and \(q/U_{MSY}\)
configurations performed similarly under the base operating model. For
the management parameters most useful in setting harvest advice,
productivity \(U_{MSY}\) and current biomass \(B_T,B_T/B_0\), the
\(q/U_{MSY}\) configuration either improved all metrics, or kept metrics
within a tolerable level of the ideal (Figure 5), e.g.~interval coverage
fell for \(U_{MSY}\), but remained within 10\% of the nominal level.
Similarly, predictive quantile \(Q(\theta)\) distributions were slightly
more uniform under the \(q/U_{MSY}\) configuration than the single-stock
model, indicating an improvement in estimator precision and bias,
however the difference between \(q\) and \(q/U_{MSY}\) configurations
was subtle. Plots of the full set of metrics for all multi-stock model
configurations and parameters under the base operating model can be
found in the supplementary material (Figures S1 - S4).

Increased precision in catchability and biomass parameters under
hierarchical multi-stock models was not always a benefit. Under a single
simulation replicate, 95\% confidence intervals of biomass estimates
from joint models were generally more precise than single-stock
estimates; however, increased precision occasionally created estimates
that were overprecise, leaving true biomass values outside confidence
intervals (Figure 6, Stock 2, \(q\) and \(Q/U_{MSY}\) models).
Furthermore, hierarchical estimation appeared to falsely detect an
increasing trend in biomass, where the single-stock model was more
conservative (Figure 5, Stock 2), but corrected the same behaviour in
the single-stock model for a different stock in the same complex (Figure
5, Stock 1).

\hypertarget{simulation-experiment-results}{%
\subsection{Simulation Experiment
Results}\label{simulation-experiment-results}}

\hypertarget{model-configuration-effects}{%
\subsubsection{Model configuration
effects}\label{model-configuration-effects}}

When comparing MARE values through the \(\Delta\) metric, multi-stock
model configurations that estimated the shared prior on survey
catchability, denoted \(q\) and \(q/U_{MSY}\), stood out as the most
beneficial for parameters of the low data quality stocks (stock
\(s = 1\)). Both of these configurations increased \(\Delta\) values, or
had effects that were within 1 standard error of zero (Table 4, Stock 1
\(\Delta\) values), indicating that multi-stock model configurations
produced MARE values at most equal to those produced by single-stock
models.

As under the base operating model, according to the \(\Delta\) metric
the best performing hierarchical multi-stock model for providing harvest
advice was \(q/U_{MSY}\). Closer inspection of \(\beta_{q}\) and
\(\beta_{U_{MSY}}\) values indicated that estimation of the mean optimal
harvest rate reduced the larger benefit to catchability in both surveys
\(q_{1,1}, q_{2,1}\) and optimal biomass \(B_{1,MSY}\) (Table 4,
\(\beta_q\) and \(\beta_{q,U_{MSY}}\)). On the other hand, while the
\(U_{MSY}\) prior had not effect on terminal biomass \((\Delta(B_T))\),
the effects on relative biomass \(\Delta(B_T/B_0)\) were nearly tripled
over the reference level \(\beta_0\). The \(\Delta\) values for optimal
biomass \(B_{MSY}\) and catchability parameters were lower, but these
parameters are not particularly critical for providing harvest advice.

The \(q\) and \(q/U_{MSY}\) configurations stood out at the complex
level also, with higher meta-model coefficients than the \(U_{MSY}\)
configuration (Table 4, Complex Aggregate \(\overline\Delta\) Values).
Under the aggregate MARE ratio \(\overline\Delta\), it was more
difficult to separate the two best models as the meta-model coefficients
for both \(q\) and \(q/U_{MSY}\) were closer together, e.g.
\(\overline\Delta(B_T)\), and there was a reduction in
\(\overline\Delta(U_{MSY})\) under the \(q/U_{MSY}\) configuration.
Unlike the stock-specific \(\Delta\) values, the prior configuration had
an effect on the \(\overline\Delta(U_{MSY})\) response in the aggregate,
where the \(q/U_{MSY}\) configuration produced the biggest reduction
\(\overline\Delta(U_{MSY})\). On the other hand, the largest increase
over the null model reference level was also produced by the
\(q/U_{MSY}\) configuration for the \(\overline\Delta(B_T/B_0)\)
response, indicating a tradeoff between estimates of stock status and
productivity.

The \(U_{MSY}\) configuration tended to perform the worst according to
the \(\Delta\) metric. We expected to see a benefit to productivity
parameter estimates but we were surprised to find there was no benefit
to a low data quality stock. Moreover, meta-model coefficients for
\(\Delta\) and \(\overline\Delta\) response variables were consistently
smaller than the other configurations, and often negative or
insignificant.

\hypertarget{factor-effects}{%
\subsubsection{Factor effects}\label{factor-effects}}

As expected, the effects of shrinkage were most beneficial under
low-information scenarios, according to the \(\Delta\) metrics. When the
biomass was initialized in a fished state, \(\Delta\) and
\(\overline\Delta\) values increased (Table 4, \(\beta_{d_{s,1}} < 0\)).
Similarly, there were significant increases in \(\Delta\) and
\(\overline\Delta\) values for all parameters when the assessments were
initialized at the beginning of survey 2 (Table 4,
\(\beta_{T_{1}} > 0\)). These improvements under low information
conditions are largely driven by a stabilising effect of shrinkage. That
is, single-stock models produced relatively larger MARE values as data
data quality was reduced. Under the same conditions, the hierarchical
multi-stock models were restricted from increasing MARE values as fast
by shrinkage (Table 4).

We found that the \(q\) and \(q/U_{MSY}\) configurations were sensitive
to data quality and the choice of performance measure. For example,
under a 1-way trip fishing history with 4 identical stocks (Figure 7),
the \(q\) configuration eliminatedd bias in \(U_{MSY}\) and improved
interval coverage from 62\% to 56\%, correcting an under-precise
estimator. In contrast, the \(q/U_{MSY}\) configuration was
over-precise, indicated by an interval coverage of 33\% and the quantile
distribution becoming slightly \(\bigcup\)-shaped, and also increased
bias in \(U_{MSY}\) estimates (Figure 7, \(U_{MSY}\)).

On the other hand, the \(q/U_{MSY}\) configuration appeared to perform
better under a 2-way trip fishing history, a short time series, and
fished initialisation. The \(q/U_{MSY}\) configuration reduced bias for
relative biomass \(B_t/B_0\) and almost eliminated bias for \(U_{MSY}\)
(Figure 8, \(U_{MSY}\)). Interval coverage also improved under the
\(q/U_{MSY}\) configuration for terminal biomass estimates \(B_T\) and
\(B_T/B_0\), coming closer to the nominal rate of 50\%. Although the
\(U_{MSY}\) interval coverage fell to 36\% under the \(q/U_{MSY}\)
configuration, indicating an over-precise estimator, we viewed this as
favourable compared to the \(q\) configuration, where \(U_{MSY}\) was
under-precise by a similar amount, yet remained positively biased.

The effect of complex size \(S\) and the number of low information
stocks \(L\) interacted in unexpected ways. According to the selected
meta-model, the size of the complex \(S\) and the number of low
information stocks \(L\) appeared to have little effect on response
values. Indeed, all \(\beta_S\) and \(\beta_L\) effects on \(\Delta\)
and \(\overline\Delta\) values were at most \(0.09\) in magnitude, if
they were included at all. These weak effects indicated that the linear
meta-model is probably too simple for these factors (Figure 9).
Increasing the number of low-information stocks \(L\) was always an
improvement for \(\Delta\) values when moving from \(L = 0\) to
\(L = 1\). This was was expected given that the \(\Delta\) values were
calculated for stock \(s = 1\) (a data poor stock if \(L > 0\)), and we
expected that multi-stock models and single-stock models would have
similar estimates when fit to complexes of data-rich stocks. Beyond
\(L = 1\) any improvements in MARE values were dependent on the size of
the complex. Generally, it appeared that keeping the number of low
information stocks under half of the complex size, i.e. \(L < S/2\),
preserved the most benefit in terms of precision, though this pattern
reversed for \(L = 3\) and \(S=4\). Complex aggregate
\(\overline\Delta\) values were comparatively flatter in response to the
levels of \(L\). We didn't produce response surfaces for other factor
combinations as these factors all had 2 levels each, meaning that a
linear model should capture the average behaviour.

\hypertarget{assessments-of-british-columbia-dover-sole}{%
\subsection{Assessments of British Columbia dover
sole}\label{assessments-of-british-columbia-dover-sole}}

Multi-stock models defined by shared catchability \(q\) and shared
catchability and optimal harvest rate configurations \(q/U_{MSY}\)
performed best for the British Columbia dover sole complex based on AICc
values. These same configurations also performed best in in the
simulation experiments. The \(U_{MSY}\) configuration and the null model
both had AICc scores more than 500 points higher than the best
performing multi-stock configuration. The selected multi-stock models
gave AICc scores between 100 and 200 units below the total single-stock
model scores under both hypotheses (Table 5, AICc), indicating that the
increase in estimated parameters was justified. All models had lower
AICc values under the assumption of low process error variance.

Hierarchical multi-stock models reduced parameter uncertainties when
compared to single-stock models. Multi-stock models with shared priors
produced lower cofficients of variation, defined as
\(CV = \sqrt{e^{se^2}-1}\), for estimates of optimal biomass and
productivity parameters, reducing coefficients of variation below 100\%
in some cases (single-stock vs multi-stock models in Table 4). Similar
reductions in uncertainty are visible in reconstructions of stock
biomass time series (Figure 10).

Assessments of the dover sole complex were qualitatively similar between
model configurations and hypotheses. The major differences between
assessment model configurations were the level of uncertainty in
parameter estimates, and the scale of each individual stock's biomass,
but the trends over time were the same (Figure 10). The Hecate Strait
(HS) stock showed increasing biomass since 1984, with more or less
process variation depending on the configuration and variance hypothesis
(Figure S5). The Queen Charlotte Sound stock showed an initial depletion
with increased landings between 2003 and 2006, followed by some growth
that has continued until present day. Finally, the West Coast of
Vancouver Island (WCVI) stock showed a flat biomass trend following
initial depletion from 2003 to 2006. The flat trend in the WCVI stock
may indicate that fishing was balancing annual production.

We found that the multi-stock assessment model configuration
\(q/U_{MSY}\) generally estimated all stocks as smaller and more
productive than other assessments (Table 5). This was most noticable for
the QCS stock biomass estimates by multi-stock models, where the
single-stock model considered the optimal biomass to be close to 18 kt,
with a terminal relative biomass between 7\% and 13\%, in contrast to
the selected multi-stock configurations, where optimal biomass was
between 3 kt and 6 kt, with a current relative biomass between 95\% and
110\%. Under the single-stock model configuration, the biomass scales
corresponded to expected catchability values of \(q_{2,HS} = 0.10\),
\(q_{2,QCS}=0.74\) and \(q_{2,WCVI} = 0.16\). We considered this
distribution of catchability values between stocks of the same species
unlikely, given that the biomass indices are relative biomass values and
catchability corresponded to trawl efficiency. It was more likely that
the single-stock assessment reduced the biomass parameter estimates for
the QCS stock because of the fished initialisation in 2003. Starting in
this state removed any depletion signal from the earlier catch history,
and allowing the model to explain the stock indices catch with a smaller
biomass.

No selected multi-stock model indicated that dover sole stocks were
overfished or experiencing overfishing, however, the uncertainty in
relative terminal biomass and harvest rate was often very high. That is,
current relative biomass estimates were always at least 60\% of
unfished, but their coefficients of variation were in some cases above
50\% of the mean estimate (Table 5). Similarly, although relative
harvest rate estimates were all at most 70\% of the optimal harvest rate
(Table 5), their coefficients of variation were at least 65\%, and
sometimes greater than 100\%, of the mean estimate for each stock under
some model configurations, most often under the high variance
assumption.

The \(q/U_{MSY}\) hierarchical multi-stock model configuration had the
best fit to the data, which is not surprising given that the dover sole
complex closely matches the scenario shown in Figure 8, with a fished
initialisation and 2 stocks having short time-series of observations.
Under those simulation experiments, the \(q/U_{MSY}\) configuration was
considered over-precise, but essentially unbiased, for \(U_{MSY}\)
estimates. In contrast, for assessments of dover sole data with low
process error variance, the precision seems be lower under the
\(q/U_{MSY}\) configuration, indicated by larger coefficients of
variation (Table 5).

\hypertarget{discussion}{%
\section{Discussion}\label{discussion}}

Our simulation results indicate that, as expected, shrinkage effects in
hierarchical multi-stock assessment models are most beneficial when some
data sets have low statistical power. Furthermore, both configurations
that estimated a shared catchability prior performed best for estimating
key management parameters. On the other hand, we found that shrinkage
does not always improve stock assessment performance relative to a
single-stock approach. In particular, the benefits of joint estimation
depend on several factors, including the information content of the
data, the choices for hierarchical model priors, and the particular
management parameters of interest.

Model configurations that shared prior distributions on survey
catchability (\(q\) and \(q/U_{MSY}\)) stood out as the best options for
improving parameter estimates for stocks with low data quality. This
result may occur because catchability is a linear parameter within the
assessment, while optimal harvest rate parameters are embedded within
non-linear popoulation dynamics. Although this hypothesis does not
explain how different configurations increase or reduce bias and
precision, it may provide a template to guide expectations and generate
hypotheses when testing other hierarchical model behaviour.

We found that simply adding a joint likelihood can have positive
effects, which was surprising because there should be no mathematical
difference between optimising a set of single-stock models independently
vs binding them in a joint model by simply adding their negative log
likelihoods together. This result may indicate a stabilising effect from
the joint likelihood, where simply including data-rich species without
shared priors improves the numerical performance of minimisation
algorithm.

There was mixed evidence that increasing the size of the assessment
complex produced better results under hierarchical multi-stock models.
For instance, in the lower information scenarios, the effect of the
complex size depended on the number of low-information stocks present in
the system. The most benefit for the first stock \(s = 1\) was realized
when moving from no low information stocks (\(L = 0\)) to one low
information stock (\(L = 1\)). This is counter-intuitive, as decreasing
information should reduce precision, but represents the stability
induced by the shrinkage from the multi-stock models. Looking at
response surfaces averaged over all factor levels and configurations, we
found that complexes of size \(S = 7\) provided the most stable benefit
(in terms of MARE values) for different numbers of low information
stocks \(L\); however, we weren't testing for an optimal size, which
would require a new design with a finer resolution on \(L\) and \(S\)
factors.

Some of our results may be caused by a discrepancy between the
underlying assumption of normality for parameter distributions used in
the Laplace approximation to the integrated likelihood and the true
parameter distribution (Kristensen et al. 2015). Despite the integrated
likelihood, the approximation by a normal distribution means that there
is potential for bias caused by disagreement between the modes of the
assumed normal distribution and true parameter distribution (Stewart et
al. 2013).

Although we investigated a single-species, multi-stock complex, where
stocks represented biologically identical management units within the
dover sole fishery, the hierarchical multi-stock approach could be
extended to a multi-species approach by simulating stocks with different
biological parameters \(B_{s,0}\) and \(r_s\). We suspect that a
differences in unfished biomass \(B_{s,0}\) would not have a strong
effect on overall performance. In a Schaefer model context, the unfished
biomass parameter determines the absolute scale at which the dynamics
operate, but has little effect on the dynamics themselves. Density
dependence in annual production is driven by this parameter, but that
effect is independent of absolute biomass and relies, instead, on the
relative biomass \(B_t/B_0\). In contrast, differences among intrinsic
growth rates may improve estimates in assessment models that estimate
shared productivity priors. More productive stocks would grow faster
when fishing pressure is reduced, reducing uncertainty in productivity
estimates for those stocks. Stocks with more precise estimates may then
have a dominating effect on the hierarchical prior, improving
hierarchical assessments but potentially biasing estimates of weaker
stock productivities (Raudenbush and Bryk 2002).

Multi-species extensions to the framework we've presented here may also
provide deeper insights. For example, introducing age-structured
population dynamics (Fournier et al. 1998), or a delay-difference
formulation (Schnute 1985), would differentiate multiple species further
than a simple Schaefer model by allowing for different maturation
delays, growth rates, and recruitment dynamics to affect stock
production. If biological data were unavailable for informing
life-history parameter estimates under more realistic population
dynamics, meta-analyses of Beverton-Holt life history invariants within
family groups could provide informative prior distributions (Nadon and
Ault 2016). Indeed, recent meta-analyses have shown that publically
available data-bases of life history parameters can be useful for this
type of application (Thorson et al. 2014). Similar meta-analyses of the
same data-bases, comparing species that are evolutionarily related,
improves the utility of life history invariants by estimating different
ratios within taxa, improving their utility as informative priors and
potentially providing inverse-gamma priors on hierarchical variance
terms in the form of evolutionary covariance estimates (Thorson et al.
2017).

We made several simplifying assumptions about the population dynamics
for simplicity in design and interpretation. In addition to assuming
that biological parameters are the same for stocks within the complex,
we assumed fishing pressure was identical among stocks, and the
magnitude of species-specific effects was identical. The choice of
identical biology removed a ``stock-effect'' on management parameter
estimates, as discussed above for productivity. With different
biological parameters, the ability to identify hierarchical estimator
effects may be reduced due to confounding with stock effects. Next,
subjecting stocks to identical fishing pressure simplifed the generation
of assessment data. Simplifying the simulations in this way may have
increased the correlation between stocks, improving performance of the
hierarchical multi-stock estimators relative to more realistic
situations. For example, it would be more realistic to link fishing
mortality to fishing effort through a stock-specific fishery
catchability.

We also made simplifying assumptions when defining the assessment model
treatment of stock-specific effect \(\zeta_{s,t}\). These assumptions
were identical standard deviations, which matched the simulated
dynamics, zero correlation in \(\zeta_{,t}\) process errors, which did
not match the simulated dynamics, and we avoided estimating the shared
year effect \(\bar\epsilon_t\), despite simulating these effects. The
reason for the second assumption was for stability in simulation trials,
as estimating the correlation often produced nonsensical results. It may
be possible to address this by applying an inverse Wishart prior for the
full estimated covariance matrix, but we did not consider this within
the scope of this research. We avoided estimating the shared year effect
as this was removed from the experimental design after it was clear that
we would be unable to reliably estimate it, and there was no benefit to
partitioning the variance across an extra process error term. Adding
another data stream, such as an environmental index (Malick et al.
2015), or forcing the year effects to resemble a periodic or trend-zero
behaviour (Walters 1986), may improve these estimates in other studies.

We did not conduct sensitivity analyses of the hyperpriors. Intuitively,
we expect that more precise inverse-gamma hyperpriors on estimated
variance parameters would increase the shrinkage effect, and thereby
clustering stock-specific estimates closer to a biased mean value.
Instead of focusing on the behaviour induced by hyperprior settings, we
chose instead to focus on the behaviour induced by defining the shared
priors, and left the hyperpriors on prior means sufficiently vague to
emulate the true prior knowledge about the dover sole complex, and on
prior variances sufficiently informative to encourage a shrinkage
effect.

Fitting the hierarchical multi-stock surplus production models
assessment to dover sole data showed that shrinkage effects carried over
to a real system. Shrinkage effects reduced uncertainty when data had
low statistical power, and provided more realistic estimates of
catchability parameters than single-stock models, especially for the
Queen Charlotte Sound stock. While the resulting estimates were
sometimes quite uncertain, and a full assessment would require more
scrutiny or a different model structure than we have provided here, our
results indicate that all three dover sole stocks are likely in a
healthy state given recent rates of exploitation.

\hypertarget{conclusion}{%
\subsection{Conclusion}\label{conclusion}}

Our results confirm that hierarchical multi-stock production models are
a feasible data-limited approach to stock assessment in multi-stock
fisheries. Under low statistical power conditions, hierarchical
multi-stock assessment modeling is preferable to data-pooling approaches
for at least two reasons. First, hierarchical multi-stock models are
able to produce stock-specific estimates that allow management decisions
to be made at a higher spatial resolution and based on data rather than
strong \emph{a priori} assumptions or management parameter values
averaged over stocks. Despite the potential for bias under low-power
conditions, stock-specific estimates of key management parameters can
provide meaningful and important feedback in the fishery management
system. Second, using a hierarchical multi-stock method ensures that an
assessment framework is readily available for more and better data,
making it much easier to update model estimates later when more data is
available. Moreover, they type of additional data to be collected could
be prioritized by examining the standard errors for observation model
components of the hierarchical multi-stock assessment models, where
higher uncertainty may indicate a better return on investments in
improved monitoring.

The feasibility of hierarchical multi-stock surplus production models
relies on catch and effort data being available, but we consider
hierarchical multi-stock production models as an important bridge
between catch-only methods and more data-intensive methods. For
instance, some catch only methods require restrictive \emph{a priori}
assumptions, such as an estimate of relative biomass as a model input
(MacCall 2009, Dick and MacCall 2011). More recently, a multi-species
assessment method was derived that removes the need for relative biomass
estimates, but requires restrictive assumptions about fishery-dependent
catchability and that all species are initially in an unfished state
(Carruthers 2018). Our approach avoids all of these assumptions. For
instance, (i) joint model estimates of relative biomass were stable in
practice, and in simulations despite absence of a current relative
biomass estimate (or assumption); (ii) hierarchical multi-stock models
have better precision when initialized in fished states; and (iii)
fishery catchability assumptions are not required. Thus, while the data
needs are higher for our approach, the potential applications are
broader in scope.

On the other hand, hierarchical multi-stock models should be scrutinized
closely via standard assessment performance measures (e.g.,
retrospective analysis) before application to real management systems.
In particular, we found that shrinkage can have unexpected non-linear
side-effects. Closed-loop simulations would be needed to determine the
long-term implications of these types of errors on multi-stock harvest
management systems (Punt et al. 2016).

\hypertarget{acknowledgements}{%
\subsection{Acknowledgements}\label{acknowledgements}}

Our funding for this research was provided by a Mitacs Cluster Grant to
S. P. Cox in collaboration with Wild Canadian Sablefish, the Pacific
Halibut Management Association and the Canadian Groundfish Research and
Conservation Society. We specifically thank A. R. Kronlund and M. Surry
at the Fisheries and Oceans Pacific Biological Station for fulfilling
data requests and helpful comments on earlier versions of the
manuscript. Further support to S.P.C. and S.D.N.J. were provided by an
NSERC Discovery Grant to S. P. Cox. We'd also like to thank the
associate editor and one anonymous reviewer for helpful comments during
the peer review of this manuscript.

\newpage

\hypertarget{references}{%
\section*{References}\label{references}}
\addcontentsline{toc}{section}{References}

\hypertarget{refs}{}
\leavevmode\hypertarget{ref-arreguin1996catchability}{}%
Arreguı'n-Sánchez, F. 1996. Catchability: A key parameter for fish stock
assessment. Reviews in fish biology and fisheries \textbf{6}(2):
221--242. Springer.

\leavevmode\hypertarget{ref-burnham2003model}{}%
Burnham, K.P., and Anderson, D.R. 2003. Model selection and multimodel
inference: A practical information-theoretic approach. Springer Science
\& Business Media.

\leavevmode\hypertarget{ref-carlin1997bayes}{}%
Carlin, B.P., and Louis, T.A. 1997. Bayes and empirical bayes methods
for data analysis. \emph{In} Statistics and Computing. Springer.

\leavevmode\hypertarget{ref-carruthers2018multispecies}{}%
Carruthers, T.R. 2018. A multispecies catch-ratio estimator of relative
stock depletion. Fisheries Research \textbf{197}: 25--33. Elsevier.

\leavevmode\hypertarget{ref-Carruthers2014Evaluating}{}%
Carruthers, T.R., Punt, A.E., Walters, C.J., MacCall, A., McAllister,
M.K., Dick, E.J., and Cope, J. 2014. Evaluating methods for setting
catch limits in data-limited fisheries. Fisheries Research
\textbf{153}(0): 48--68. doi:
\href{http://dx.doi.org/10.1016/j.fishres.2013.12.014}{http://dx.doi.org/10.1016/j.fishres.2013.12.014}.

\leavevmode\hypertarget{ref-cox2011management}{}%
Cox, S., Kronlund, A., and Lacko, L. 2011. Management procedures for the
multi-gear sablefish (anoplopoma fimbria) fishery in british columbia,
canada. Can. Sci. Advis. Secret. Res. Doc \textbf{62}.

\leavevmode\hypertarget{ref-dick2011depletion}{}%
Dick, E., and MacCall, A.D. 2011. Depletion-based stock reduction
analysis: A catch-based method for determining sustainable yields for
data-poor fish stocks. Fisheries Research \textbf{110}(2): 331--341.
Elsevier.

\leavevmode\hypertarget{ref-fargo1998flatfish-s}{}%
Fargo, J. 1998. Flatfish stock assessments for the west coast of Canada
for 1997 and recommended yield options for 1998. Canadian Stock
Assessment Secretariat Research Document.

\leavevmode\hypertarget{ref-fargo1999flatfish-s}{}%
Fargo, J. 1999. Flatfish stock assessments for the west coast of Canada
for 1999 and recommended yield options for 2000. DFO Can. Stock. Assess.
Sec. Res. Doc. (1999/199): 51.

\leavevmode\hypertarget{ref-fournier1998multifan}{}%
Fournier, D.A., Hampton, J., and Sibert, J.R. 1998. MULTIFAN-cl: A
length-based, age-structured model for fisheries stock assessment, with
application to south pacific albacore, thunnus alalunga. Canadian
Journal of Fisheries and Aquatic Sciences \textbf{55}(9): 2105--2116.
NRC Research Press.

\leavevmode\hypertarget{ref-gelman2014bayesian}{}%
Gelman, A., Carlin, J.B., Stern, H.S., and Rubin, D.B. 2014. Bayesian
data analysis. Taylor \& Francis.

\leavevmode\hypertarget{ref-hilborn1979comparison}{}%
Hilborn, R. 1979. Comparison of fisheries control systems that utilize
catch and effort data. Journal of the Fisheries Board of Canada
\textbf{36}(12): 1477--1489. NRC Research Press.

\leavevmode\hypertarget{ref-hilborn1992quantitative}{}%
Hilborn, R., and Walters, C.J. 1992. Quantitative fisheries stock
assessment: Choice, dynamics and uncertainty/book and disk. Springer
Science \& Business Media.

\leavevmode\hypertarget{ref-james1961estimation}{}%
James, W., and Stein, C. 1961. Estimation with quadratic loss. \emph{In}
Proceedings of the fourth berkeley symposium on mathematical statistics
and probability. pp. 361--379.

\leavevmode\hypertarget{ref-jiao2011poor}{}%
Jiao, Y., Cortés, E., Andrews, K., and Guo, F. 2011. Poor-data and
data-poor species stock assessment using a bayesian hierarchical
approach. Ecological Applications \textbf{21}(7): 2691--2708. Wiley
Online Library.

\leavevmode\hypertarget{ref-jiao2009hierarchical}{}%
Jiao, Y., Hayes, C., and Cortés, E. 2009. Hierarchical Bayesian approach
for population dynamics modelling of fish complexes without
species-specific data. ICES Journal of Marine Science: Journal du
Conseil \textbf{66}(2): 367--377. Oxford University Press.

\leavevmode\hypertarget{ref-kleijnen2008design}{}%
Kleijnen, J.P. 2008. Design and analysis of simulation experiments.
Springer.

\leavevmode\hypertarget{ref-kristensen2015tmb}{}%
Kristensen, K., Nielsen, A., Berg, C.W., Skaug, H., and Bell, B. 2015.
TMB: Automatic differentiation and laplace approximation. arXiv preprint
arXiv:1509.00660.

\leavevmode\hypertarget{ref-maccall2009depletion}{}%
MacCall, A.D. 2009. Depletion-corrected average catch: A simple formula
for estimating sustainable yields in data-poor situations. ICES Journal
of Marine Science: Journal du Conseil \textbf{66}(10): 2267--2271.
Oxford University Press.

\leavevmode\hypertarget{ref-magnusson2007makes}{}%
Magnusson, A., and Hilborn, R. 2007. What makes fisheries data
informative? Fish and Fisheries \textbf{8}(4): 337--358. Wiley Online
Library.

\leavevmode\hypertarget{ref-malick2015accounting}{}%
Malick, M.J., Cox, S.P., Peterman, R.M., Wainwright, T.C., Peterson,
W.T., and Krkošek, M. 2015. Accounting for multiple pathways in the
connections among climate variability, ocean processes, and coho salmon
recruitment in the northern california current. Canadian Journal of
Fisheries and Aquatic Sciences \textbf{72}(10): 1552--1564. NRC Research
Press.

\leavevmode\hypertarget{ref-mccullagh1984generalized}{}%
McCullagh, P. 1984. Generalized linear models. European Journal of
Operational Research \textbf{16}(3): 285--292. Elsevier.

\leavevmode\hypertarget{ref-mueter2002spatial}{}%
Mueter, F.J., Ware, D.M., and Peterman, R.M. 2002. Spatial correlation
patterns in coastal environmental variables and survival rates of salmon
in the north-east pacific ocean. Fisheries Oceanography \textbf{11}(4):
205--218. Wiley Online Library.

\leavevmode\hypertarget{ref-nadon2016stepwise}{}%
Nadon, M.O., and Ault, J.S. 2016. A stepwise stochastic simulation
approach to estimate life history parameters for data-poor fisheries.
Canadian Journal of Fisheries and Aquatic Sciences \textbf{73}(12):
1874--1884. NRC Research Press.

\leavevmode\hypertarget{ref-peterman1990statistical}{}%
Peterman, R.M. 1990. Statistical power analysis can improve fisheries
research and management. Canadian Journal of Fisheries and Aquatic
Sciences \textbf{47}(1): 2--15. NRC Research Press.

\leavevmode\hypertarget{ref-peterman1998patterns}{}%
Peterman, R.M., Pyper, B.J., Lapointe, M.F., Adkison, M.D., and Walters,
C.J. 1998. Patterns of covariation in survival rates of british
columbian and alaskan sockeye salmon (oncorhynchus nerka) stocks.
Canadian Journal of Fisheries and Aquatic Sciences \textbf{55}(11):
2503--2517. NRC Research Press.

\leavevmode\hypertarget{ref-punt2003extending}{}%
Punt, A.E. 2003. Extending production models to include process error in
the population dynamics. Canadian Journal of Fisheries and Aquatic
Sciences \textbf{60}(10): 1217--1228. NRC Research Press.

\leavevmode\hypertarget{ref-punt2016management}{}%
Punt, A.E., Butterworth, D.S., Moor, C.L., De Oliveira, J.A., and
Haddon, M. 2016. Management strategy evaluation: Best practices. Fish
and Fisheries. Wiley Online Library.

\leavevmode\hypertarget{ref-punt2002evaluation}{}%
Punt, A.E., Smith, A.D., and Cui, G. 2002. Evaluation of management
tools for australia's south east fishery. 1. Modelling the south east
fishery taking account of technical interactions. Marine and Freshwater
Research \textbf{53}(3): 615--629. CSIRO.

\leavevmode\hypertarget{ref-punt2005using}{}%
Punt, A.E., Smith, D.C., and Koopman, M.T. 2005. Using information for
data-rich species to inform assessments of data-poor species through
bayesian stock assessment methods. Primary Industries Research Victoria.

\leavevmode\hypertarget{ref-punt2011among}{}%
Punt, A.E., Smith, D.C., and Smith, A.D. 2011. Among-stock comparisons
for improving stock assessments of data-poor stocks: The ``Robin Hood''
approach. ICES Journal of Marine Science: Journal du Conseil
\textbf{68}(5): 972--981. Oxford University Press.

\leavevmode\hypertarget{ref-raudenbush2002hierarchical}{}%
Raudenbush, S.W., and Bryk, A.S. 2002. Hierarchical linear models:
Applications and data analysis methods. Sage.

\leavevmode\hypertarget{ref-rproject2015}{}%
R Core Team. 2015. R: A language and environment for statistical
computing. R Foundation for Statistical Computing, Vienna, Austria.
Available from \url{http://www.R-project.org/}.

\leavevmode\hypertarget{ref-schnute1985general}{}%
Schnute, J. 1985. A general theory for analysis of catch and effort
data. Canadian Journal of Fisheries and Aquatic Sciences \textbf{42}(3):
414--429. NRC Research Press.

\leavevmode\hypertarget{ref-schnute1995influence}{}%
Schnute, J.T., and Richards, L.J. 1995. The influence of error on
population estimates from catch-age models. Canadian Journal of
Fisheries and Aquatic Sciences \textbf{52}(10): 2063--2077. NRC Research
Press.

\leavevmode\hypertarget{ref-stewart2013comparison}{}%
Stewart, I.J., Hicks, A.C., Taylor, I.G., Thorson, J.T., Wetzel, C., and
Kupschus, S. 2013. A comparison of stock assessment uncertainty
estimates using maximum likelihood and bayesian methods implemented with
the same model framework. Fisheries Research \textbf{142}: 37--46.
Elsevier.

\leavevmode\hypertarget{ref-thorson2015giants}{}%
Thorson, J.T., Cope, J.M., Kleisner, K.M., Samhouri, J.F., Shelton,
A.O., and Ward, E.J. 2015. Giants' shoulders 15 years later: Lessons,
challenges and guidelines in fisheries meta-analysis. Fish and Fisheries
\textbf{16}(2): 342--361. Wiley Online Library.

\leavevmode\hypertarget{ref-thorson2014assessing}{}%
Thorson, J.T., Cope, J.M., and Patrick, W.S. 2014. Assessing the quality
of life history information in publicly available databases. Ecological
Applications \textbf{24}(1): 217--226. Wiley Online Library.

\leavevmode\hypertarget{ref-thorson2017predicting}{}%
Thorson, J.T., Munch, S.B., Cope, J.M., and Gao, J. 2017. Predicting
life history parameters for all fishes worldwide. Ecological
Applications \textbf{27}(8): 2262--2276. Wiley Online Library.

\leavevmode\hypertarget{ref-walters1986adaptive}{}%
Walters, C. 1986. Adaptive management of renewable resources. MacMillan
Pub. Co., New York, NY.

\newpage
\singlespacing

\newpage

\hypertarget{tables}{%
\section{Tables}\label{tables}}

\listoftables

\newpage

\rowcolors{2}{gray!6}{white}
\begin{table}[!h]

\caption{\label{tab:simTable}Operating model parameters and their values}
\centering
\resizebox{\linewidth}{!}{
\begin{tabular}[t]{lcc}
\hiderowcolors
\toprule
Description & Symbol & Value\\
\midrule
\showrowcolors
Unfished Biomass & $B_{s,o}$ & 40kt\\
Intrinsic Rate of Growth & $r_s$ & 0.16\\
Shared Process Error SD & $\kappa$ & 0.071\\
Stock-specific Process Error SD & $\sigma_s$ & 0.071\\
Simulation Historical Period & $(T_{init},\dots,T)$ & $(1984,\dots,2016)$\\
\bottomrule
\end{tabular}}
\end{table}
\rowcolors{2}{white}{white}

\newpage

\rowcolors{2}{gray!6}{white}
\begin{table}[!h]

\caption{\label{tab:expTable}Experimental factors and their levels}
\centering
\resizebox{\linewidth}{!}{
\begin{tabular}[t]{l>{\centering\arraybackslash}p{5cm}>{\centering\arraybackslash}p{5cm}}
\hiderowcolors
\toprule
Description & Levels & Notes\\
\midrule
\showrowcolors
Fishing History & 1-way, 2-way trips & Low/High contrast in biomass\\
Complex Size, $S$ & 4,7,10 & \\
Low data quality stocks, $L$ & 0,1,2,3 & \\
Initial Assessment Year 1984, 2003 & Short or long series of observations ($t = 1$ or $t = 20$ of $T=34$ years) & \\
Initial Relative Depletion & 0.4, 0.7, 1.0 & Fished or unfished initialisation\\
\bottomrule
\end{tabular}}
\end{table}
\rowcolors{2}{white}{white}

\newpage

\rowcolors{2}{gray!6}{white}
\begin{table}[!h]

\caption{\label{tab:priorTable}Multi- and single level priors used in the assessment model.}
\centering
\begin{tabular}[t]{cl}
\hiderowcolors
\toprule
No. & Distribution\\
\midrule
\showrowcolors
\addlinespace[0.3em]
\multicolumn{2}{l}{\textbf{Survey Catchability}}\\
\addlinespace[0.3em]
\multicolumn{2}{l}{\textit{Multi-level prior}}\\
\hspace{1em}\hspace{1em}q.1 & $\hat{q}_{o,s} \sim \log N(\log\hat{\bar{q}}_o,\hat{\iota}_{o})$\\
\hspace{1em}\hspace{1em}q.2 & $\hat{\bar{q}}_o \sim N(m_q,s_q)$\\
\hspace{1em}\hspace{1em}q.3 & $\hat{\iota}_o^2 \sim IG(\alpha_q,\beta_q)$\\
\addlinespace[0.3em]
\multicolumn{2}{l}{\textit{Single level prior}}\\
\hspace{1em}\hspace{1em}q.4 & $\hat{q}_{o,s}\sim N(m_q,s_q)$\\
\addlinespace[0.3em]
\multicolumn{2}{l}{\textbf{Optimal Harvest Rate}}\\
\addlinespace[0.3em]
\multicolumn{2}{l}{\textit{Multi-level prior}}\\
\hspace{1em}\hspace{1em}U.1 & $\hat{U}_{s,MSY} \sim \log N(\log\hat{\bar{U}}_{MSY},\hat{\sigma}_{U})$\\
\hspace{1em}\hspace{1em}U.2 & $\hat{\bar{U}}_{MSY} \sim N(m_U, s_U)$\\
\hspace{1em}\hspace{1em}U.3 & $\hat{\sigma}_U^2 \sim IG(\alpha_U,\beta_U)$\\
\addlinespace[0.3em]
\multicolumn{2}{l}{\textit{Single level prior}}\\
\hspace{1em}\hspace{1em}U.4 & $\hat{U}_{s,MSY} \sim N(m_U,s_U)$\\
\bottomrule
\end{tabular}
\end{table}
\rowcolors{2}{white}{white}

\newpage

\begin{landscape}\rowcolors{2}{gray!6}{white}
\begin{table}

\caption{\label{tab:perfTable}Meta-model coefficients for multi-stock assessment model prior configurations (columns 3-5) and experimental factors (cols 6-10). Response variables are $\Delta(\theta_s) = \frac{MARE_{MS}(\theta_s)}{MARE_{SS}(\theta_s)} - 1$ values for stock $s=1$ (rows 1-6), complex aggregate $\overline\Delta(\theta) = \frac{\sum_s MARE_{MS}(\theta_s)}{\sum_s MARE_{SS}(\theta_s)} - 1$ values (rows 7-12), single stock assessment MARE values for stock 1 (rows 13-18), and multi-stock model MARE values for stock 1 (rows 19 - 24). The intercept (col 2) is the average value of the response across all factors, and represents the null model configuration in rows 1-12 and 19-24. Coefficients of multi-stock model prior configurations independently give the average contribution of that configuration to the response value, while coefficients for experimental factors are calculated based on rescaling factors to the interval $[-1,1]$. This means the contribution of each factor to the response is equal to its coefficient at the maximum factor value, and the negative value of its coefficient at the minimum factor value. Response values are found by summing across the rows, \textit{taking only one prior configuration coefficient}, and scaling factor coefficients as necessary.}
\centering
\resizebox{\linewidth}{!}{
\begin{tabular}[t]{cccccccccc}
\hiderowcolors
\toprule
\multicolumn{2}{c}{\bfseries  } & \multicolumn{3}{c}{\bfseries Prior Configuration} & \multicolumn{5}{c}{\bfseries Experimental Factor} \\
\cmidrule(l{2pt}r{2pt}){3-5} \cmidrule(l{2pt}r{2pt}){6-10}
\multicolumn{1}{c}{Response} & \multicolumn{1}{c}{Ref Level} & \multicolumn{1}{c}{ } & \multicolumn{1}{c}{ } & \multicolumn{1}{c}{ } & \multicolumn{1}{c}{Init. Dep} & \multicolumn{1}{c}{Init. Assessment} & \multicolumn{1}{c}{Low Data Stocks} & \multicolumn{1}{c}{Complex Size} & \multicolumn{1}{c}{Fishing History} \\
\cmidrule(l{2pt}r{2pt}){1-1} \cmidrule(l{2pt}r{2pt}){2-2} \cmidrule(l{2pt}r{2pt}){6-6} \cmidrule(l{2pt}r{2pt}){7-7} \cmidrule(l{2pt}r{2pt}){8-8} \cmidrule(l{2pt}r{2pt}){9-9} \cmidrule(l{2pt}r{2pt}){10-10}
  & $\beta_0$ & $\beta_q$ & $\beta_{U_{MSY}}$ & $\beta_{q/U_{MSY}}$ & $\beta_{d_{1,1}}$ & $\beta_{T_1}$ & $\beta_L$ & $\beta_S$ & $\beta_U$\\
\midrule
\showrowcolors
\addlinespace[0.3em]
\multicolumn{10}{l}{\textbf{Low Data Quality Stock ($s = 1$) $\Delta$ Values}}\\
\hspace{1em}$\Delta(U_{1,MSY})$ & 0.60 (0.07) & 0.25 (0.09) & 0.04 (0.09) & 0.09 (0.09) & -0.14 (0.04) & 0.35 (0.04) & - & - & -\\
\hspace{1em}$\Delta(B_{1,T})$ & -0.01 (0.04) & 0.28 (0.05) & -0.02 (0.05) & 0.28 (0.05) & -0.04 (0.02) & 0.08 (0.02) & - & - & 0.11 (0.02)\\
\hspace{1em}$\Delta(B_{1,MSY})$ & 0.16 (0.03) & 0.10 (0.04) & -0.07 (0.04) & 0.02 (0.04) & -0.06 (0.02) & 0.11 (0.02) & 0.06 (0.02) & - & -0.02 (0.01)\\
\hspace{1em}$\Delta(B_{1,T}/B_{1,0})$ & 0.32 (0.07) & 0.29 (0.10) & 0.13 (0.10) & 0.63 (0.10) & -0.09 (0.04) & 0.30 (0.04) & - & 0.09 (0.04) & 0.14 (0.03)\\
\hspace{1em}$\Delta(q_{1,1})$ & 0.06 (0.05) & 0.46 (0.07) & -0.01 (0.07) & 0.27 (0.07) & - & 0.15 (0.03) & - & - & 0.06 (0.03)\\
\hspace{1em}$\Delta(q_{2,1})$ & -0.02 (0.02) & 0.23 (0.03) & -0.05 (0.03) & 0.10 (0.03) & - & - & 0.03 (0.02) & -0.04 (0.01) & 0.08 (0.01)\\
\addlinespace[0.3em]
\multicolumn{10}{l}{\textbf{Complex Aggregate $\overline\Delta$ Values}}\\
\hspace{1em}$\overline\Delta(U_{MSY})$ & 0.47 (0.04) & 0.13 (0.04) & -0.07 (0.04) & -0.07 (0.04) & -0.10 (0.03) & 0.21 (0.03) & 0.04 (0.03) & - & -0.03 (0.02)\\
\hspace{1em}$\overline\Delta(B_T)$ & 0.04 (0.02) & 0.22 (0.02) & -0.03 (0.02) & 0.21 (0.02) & -0.05 (0.01) & 0.11 (0.01) & -0.03 (0.01) & 0.02 (0.01) & 0.07 (0.01)\\
\hspace{1em}$\overline\Delta(B_{MSY})$ & 0.11 (0.02) & 0.11 (0.02) & -0.05 (0.02) & 0.01 (0.02) & -0.07 (0.01) & 0.09 (0.02) & - & - & 0.03 (0.01)\\
\hspace{1em}$\overline\Delta(B_T/B_0)$ & 0.31 (0.04) & 0.25 (0.04) & 0.03 (0.04) & 0.39 (0.04) & -0.08 (0.02) & 0.24 (0.03) & - & - & 0.06 (0.01)\\
\hspace{1em}$\overline\Delta(q_1)$ & 0.08 (0.03) & 0.31 (0.03) & -0.02 (0.03) & 0.21 (0.03) & -0.05 (0.02) & 0.15 (0.02) & - & - & 0.08 (0.01)\\
\hspace{1em}$\overline\Delta(q_2)$ & -0.06 (0.02) & 0.26 (0.02) & -0.04 (0.02) & 0.15 (0.02) & -0.03 (0.01) & - & - & -0.02 (0.01) & 0.07 (0.01)\\
\addlinespace[0.3em]
\multicolumn{10}{l}{\textbf{Single-Stock Assessment MARE values}}\\
\hspace{1em}$U_{1,MSY}$ & 40.52 (1.22) & - & - & - & -6.64 (1.49) & 4.44 (1.21) & 3.90 (1.66) & - & -9.00 (1.08)\\
\hspace{1em}$B_{1,T}$ & 29.01 (0.56) & - & - & - & -0.96 (0.64) & 2.62 (0.54) & - & 1.01 (0.62) & 2.65 (0.51)\\
\hspace{1em}$B_{1,MSY}$ & 26.61 (0.49) & - & - & - & -5.56 (0.60) & 3.67 (0.49) & 3.11 (0.67) & -0.77 (0.54) & -\\
\hspace{1em}$B_{1,T}/B_{1,0}$ & 56.13 (1.97) & - & - & - & -11.68 (2.28) & 17.71 (1.93) & - & - & 14.34 (1.81)\\
\hspace{1em}$q_{1,1}$ & 19.58 (0.44) & - & - & - & - & 3.46 (0.44) & - & - & -\\
\hspace{1em}$q_{2,1}$ & 17.97 (0.41) & - & - & - & - & 0.59 (0.41) & - & -0.94 (0.49) & -1.00 (0.40)\\
\addlinespace[0.3em]
\multicolumn{10}{l}{\textbf{Multi-Stock Assessment MARE values}}\\
\hspace{1em}$U_{MSY}$ & 24.96 (0.87) & -3.54 (1.20) & 0.55 (1.20) & -0.13 (1.20) & -1.70 (0.59) & -0.88 (0.47) & 1.13 (0.65) & -0.78 (0.52) & -5.14 (0.42)\\
\hspace{1em}$B_T$ & 29.10 (0.76) & -6.22 (1.06) & 0.67 (1.06) & -5.80 (1.06) & - & 0.64 (0.39) & - & 0.76 (0.46) & -\\
\hspace{1em}$B_{MSY}$ & 22.85 (0.78) & -1.85 (1.06) & 1.28 (1.06) & -0.32 (1.06) & -4.23 (0.52) & 1.28 (0.42) & 1.90 (0.58) & - & -\\
\hspace{1em}$B_T/B_0$ & 40.65 (1.87) & -8.77 (2.56) & -5.02 (2.56) & -13.88 (2.56) & -5.47 (1.26) & 4.24 (1.02) & 2.49 (1.40) & -1.67 (1.12) & 6.22 (0.91)\\
\hspace{1em}$q_{1}$ & 18.51 (0.75) & -5.39 (1.04) & -0.06 (1.04) & -3.33 (1.04) & 0.69 (0.46) & 1.64 (0.39) & - & - & -1.37 (0.37)\\
\hspace{1em}$q_{1}$ & 18.51 (0.81) & -3.73 (1.15) & 1.24 (1.15) & -1.54 (1.15) & - & - & - & - & -2.58 (0.41)\\
\bottomrule
\end{tabular}}
\end{table}
\rowcolors{2}{white}{white}
\end{landscape}

\newpage

\begin{landscape}\rowcolors{2}{gray!6}{white}
\begin{table}

\caption{\label{tab:fitTable}Selected management parameter mean estimates, their coefficients of variation in parentheses, and corrected Akaike’s Information Criterion (AICc) values for selected stock assessments applied to the real dover sole data under the High and Low process error variance hypotheses. Model labels for multi-stock models indicate the shared priors used in the fitting process. Total AICc values for the Single-Stock model are given for direct comparison with the multi-stock models.}
\centering
\resizebox{\linewidth}{!}{
\begin{tabular}[t]{cccccccccc}
\hiderowcolors
\toprule
\multicolumn{1}{c}{\bfseries  } & \multicolumn{3}{c}{\bfseries High Process Error Variance} & \multicolumn{2}{c}{\bfseries  } & \multicolumn{3}{c}{\bfseries Low Process Error Variance} & \multicolumn{1}{c}{\bfseries  } \\
\cmidrule(l{2pt}r{2pt}){2-4} \cmidrule(l{2pt}r{2pt}){7-9}
Model Config & HS & QCS & WCVI & Total & Model Config & HS & QCS & WCVI & Total\\
\midrule
\showrowcolors
\addlinespace[0.3em]
\multicolumn{10}{l}{\textbf{$U_{MSY}$}}\\
\hspace{1em}Single-Stock & 0.147 (0.69) & 0.066 (1.14) & 0.122 (0.94) & - & Single-Stock & 0.113 (0.77) & 0.100 (0.71) & 0.136 (0.84) & -\\
\hspace{1em}$q$ & 0.127 (0.64) & 0.092 (0.88) & 0.095 (0.90) & - & $q$ & 0.115 (0.63) & 0.097 (0.83) & 0.104 (0.83) & -\\
\hspace{1em}$q/U_{MSY}$ & 0.205 (0.64) & 0.191 (0.73) & 0.214 (0.78) & - & $q/U_{MSY}$ & 0.156 (0.76) & 0.151 (0.74) & 0.170 (0.86) & -\\
\addlinespace[0.3em]
\multicolumn{10}{l}{\textbf{$B_T$}}\\
\hspace{1em}Single-Stock & 33.189 (1.04) & 4.841 (1.27) & 11.487 (0.89) & - & Single-Stock & 29.641 (1.13) & 2.868 (0.66) & 10.956 (0.83) & -\\
\hspace{1em}$q$ & 27.112 (0.82) & 13.843 (0.85) & 13.616 (0.74) & - & $q$ & 25.498 (0.79) & 9.873 (0.87) & 11.618 (0.77) & -\\
\hspace{1em}$q/U_{MSY}$ & 21.067 (0.91) & 11.246 (0.93) & 11.685 (0.83) & - & $q/U_{MSY}$ & 18.124 (0.96) & 7.553 (1.05) & 9.457 (0.88) & -\\
\addlinespace[0.3em]
\multicolumn{10}{l}{\textbf{$B_T/B_0$}}\\
\hspace{1em}Single-Stock & 0.968 (0.67) & 0.131 (1.97) & 0.718 (0.80) & - & Single-Stock & 0.874 (0.48) & 0.077 (1.53) & 0.700 (0.59) & -\\
\hspace{1em}$q$ & 0.917 (0.62) & 1.091 (0.73) & 0.702 (0.92) & - & $q$ & 0.842 (0.47) & 0.950 (0.47) & 0.631 (0.70) & -\\
\hspace{1em}$q/U_{MSY}$ & 0.932 (0.57) & 1.071 (0.56) & 0.830 (0.52) & - & $q/U_{MSY}$ & 0.878 (0.41) & 0.967 (0.41) & 0.708 (0.54) & -\\
\addlinespace[0.3em]
\multicolumn{10}{l}{\textbf{$U_T/U_{MSY}$}}\\
\hspace{1em}Single-Stock & 0.081 (1.03) & 0.597 (1.12) & 0.520 (1.11) & - & Single-Stock & 0.117 (0.83) & 0.669 (0.65) & 0.488 (0.96) & -\\
\hspace{1em}$q$ & 0.114 (0.85) & 0.151 (1.07) & 0.560 (1.04) & - & $q$ & 0.134 (0.67) & 0.199 (1.03) & 0.602 (0.97) & -\\
\hspace{1em}$q/U_{MSY}$ & 0.091 (0.87) & 0.089 (1.00) & 0.291 (0.95) & - & $q/U_{MSY}$ & 0.139 (0.66) & 0.168 (1.01) & 0.453 (0.88) & -\\
\addlinespace[0.3em]
\multicolumn{10}{l}{\textbf{$B_{MSY}$}}\\
\hspace{1em}Single-Stock & 17.143 (0.90) & 18.415 (1.53) & 8.004 (0.88) & - & Single-Stock & 16.951 (1.11) & 18.672 (1.48) & 7.830 (0.72) & -\\
\hspace{1em}$q$ & 14.790 (0.73) & 6.343 (0.91) & 9.693 (0.95) & - & $q$ & 15.142 (0.80) & 5.196 (0.81) & 9.210 (0.72) & -\\
\hspace{1em}$q/U_{MSY}$ & 11.306 (0.83) & 5.250 (0.86) & 7.043 (0.75) & - & $q/U_{MSY}$ & 10.323 (0.94) & 3.904 (0.97) & 6.680 (0.78) & -\\
\addlinespace[0.3em]
\multicolumn{10}{l}{\textbf{$AICc$}}\\
\hspace{1em}Single-Stock & -102.06 & -22.487 & -23.655 & -148.202 & Single-Stock & -163.261 & -54.035 & -56.802 & -274.098\\
\hspace{1em}$q$ &  &  &  & -169.838 & $q$ &  &  &  & -343.312\\
\hspace{1em}$q/U_{MSY}$ &  &  &  & -252.292 & $q/U_{MSY}$ &  &  &  & -417.676\\
\bottomrule
\end{tabular}}
\end{table}
\rowcolors{2}{white}{white}
\end{landscape}

\newpage

\hypertarget{figures}{%
\section{Figures}\label{figures}}

\listoffigures

\newpage

\begin{figure}[p]

{\centering \includegraphics[width=0.9\linewidth]{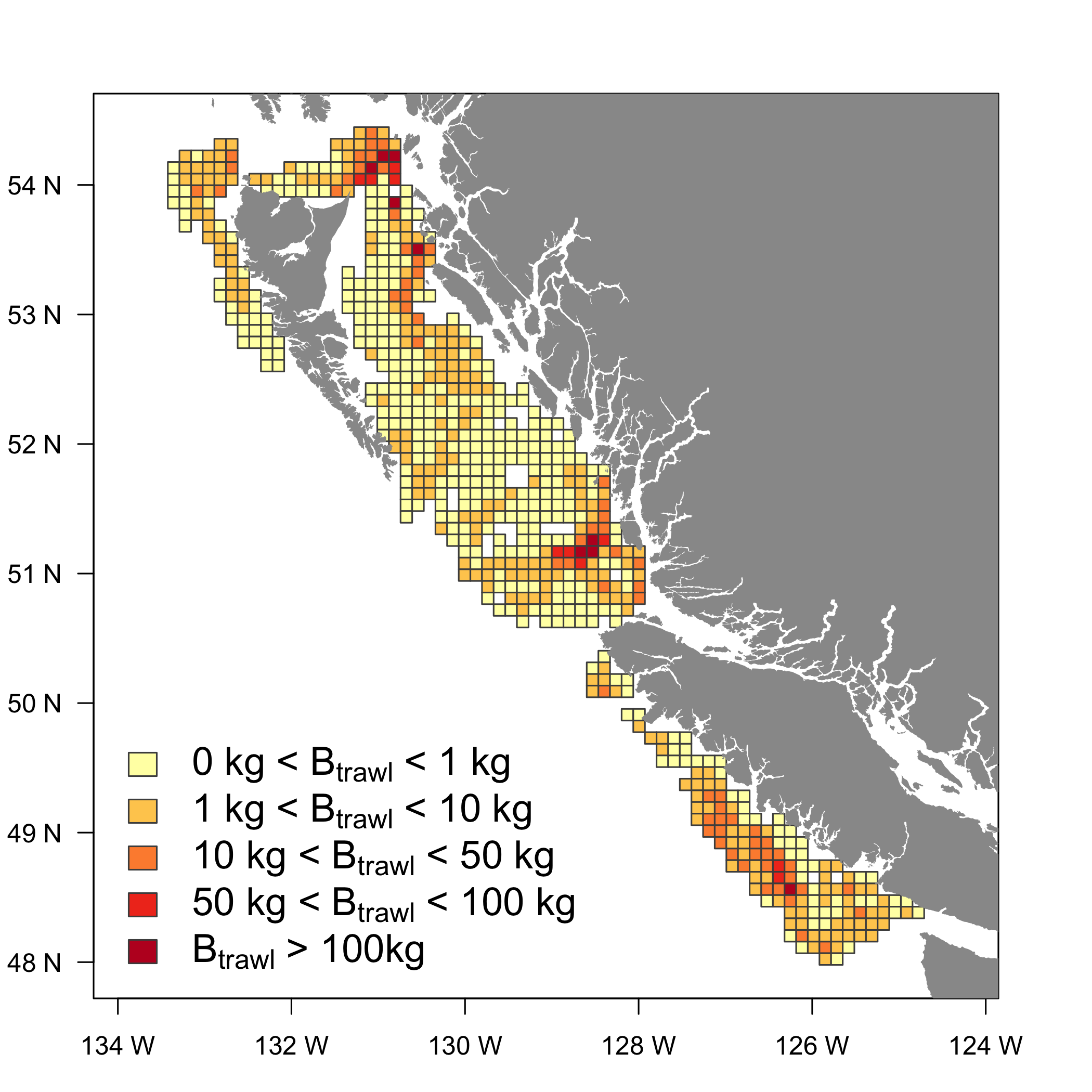} 

}

\caption{Mininum trawlable biomass $B_{trawl}$ estimates for Dover Sole on the BC coast, aggregated to a 10km square grid. Estimates are produced by scaling average trawl survey ($kg/m^2$) density values in each grid cell by the cell's area in $m^2$. Locations that do not show a coloured grid cell do not have any survey blocks from which to calculate relative biomass. Survey density data is taken from the GFBio data base maintained at the Pacific Biological Station of Fisheries and Oceans, Canada.}\label{fig:heatMapPlot}
\end{figure}

\newpage

\begin{figure}[p]

{\centering \includegraphics[width=0.9\linewidth]{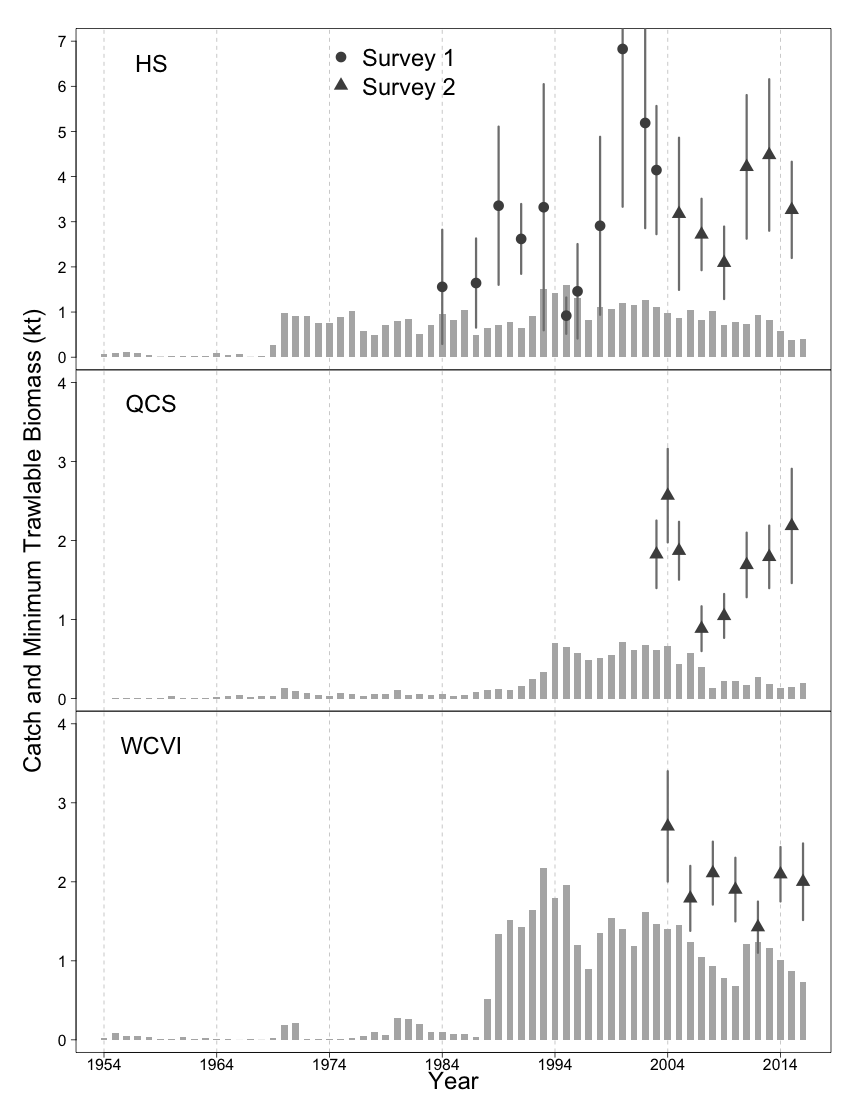} 

}

\caption{Time series of coastwide catch since 1954 (vertical bars) and relative biomass since 1984 (data points) for the three Dover Sole stocks: Haida Gwaii (HG), Queen Charlotte Sound (QCS) and West Coast of Vancouver Island (WCVI). The catch data are taken from the GFcatch, PacHarvTrawl and GFFOS data bases and trawl survey data were obtained from the GFBIO data base. All data bases are maintained at the Pacific Biological Station of Fisheries and Oceans, Canada.}\label{fig:doverDataPlot}
\end{figure}
\newpage

\begin{landscape}

\begin{figure}[p]

{\centering \includegraphics[width=0.9\linewidth]{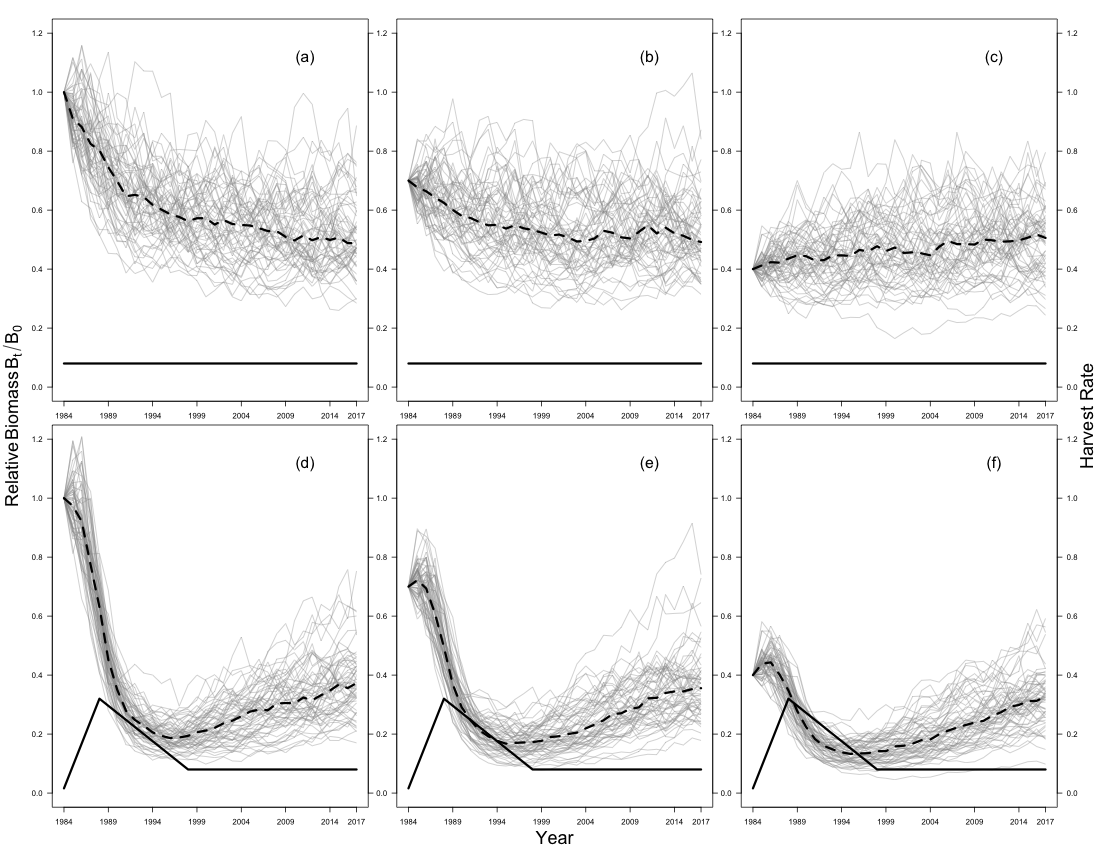} 

}

\caption{Biomass depletion trajectories of 60 random replicates under different historical fishing intensities and initial relative biomass. Plots (a) - (c) show the constant optimal harvest rate fishing history, which result in more one-way trips, and plots (d) - (f) show the two-way trip fishing history. Initial relative biomass of 40\% (panels (a), (d)), 70\%  (panels (b), (e)), and 100\% (panels (c), (f)) of $B_0$ are shown. The grey lines are traces from selected replicates, while the black dashed line is the median time series for those replicates, and the solid black line is the simulated harvest rate.}\label{fig:depPlot}
\end{figure}

\end{landscape}

\newpage

\begin{figure}[p]

{\centering \includegraphics[width=0.6\linewidth]{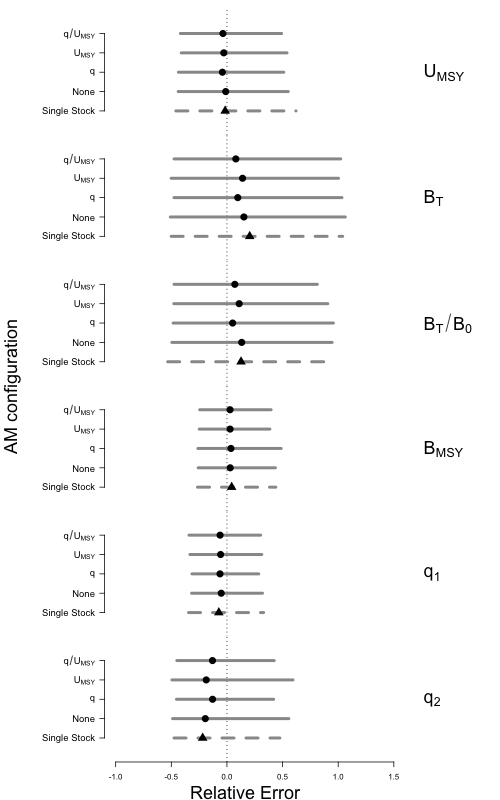} 

}

\caption{Relative error distributions for stock $s = 1$ leading and derived parameters estimated by the single stock (dashed lines and triangular points) and 4 multi-stock assessment models (solid lines and circular points) fit to data from the base operating model. Points indicate median relative errors and the grey lines the central 95\% of the relative error distribution. From the top, parameters are optimal exploitation rate ($U_{MSY}$), terminal biomass ($B_{T}$), optimal equilibrium biomass ($B_{MSY}$), terminal biomass relative to unfished ($B_{T}/B_{0}$), and catchability from surveys 1 ($q_{1}$) and 2 ($q_{2}$). Assessment model (AM) configurations indicate the single stock model, or the parameters that had hierarchical prior distribution hyperparameters estimated in the multi-stock assessment model (e.g, $q/U_{MSY}$ indicates that shared priors on both catchability and productivity were estimated).}\label{fig:mreDists}
\end{figure}

\newpage

\begin{landscape}

\begin{figure}[p]

{\centering \includegraphics[width=0.9\linewidth]{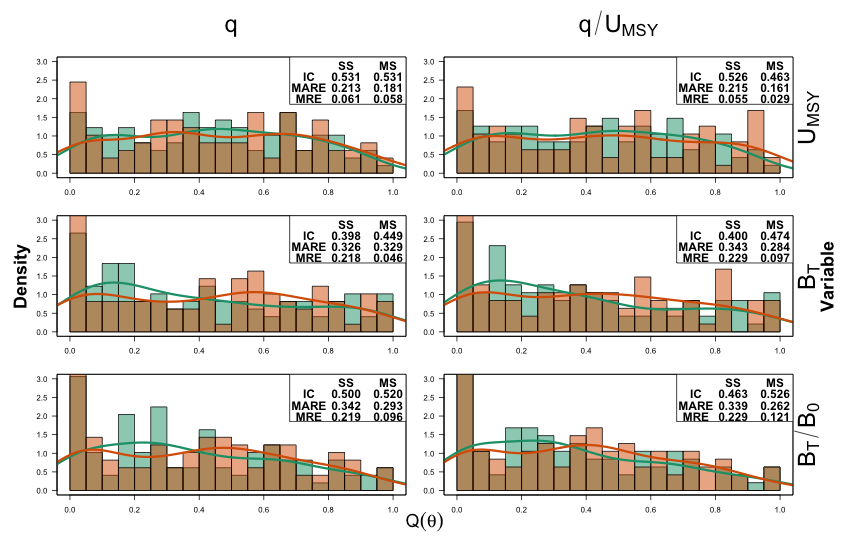} 

}

\caption{Density of predictive quantiles $Q(\theta)$ for estimates of key management parameters (rows) from single stock and $q$ and $q/U_{MSY}$ hierarchical multi-stock model configuration under the base operating model. Bars show probability density  of $Q$ distributions, with lines showing the kernel smoothed density for easier comparison between single stock (green) and multi-stock (orange) models. Top right hand corners of each panel show interval coverage (IC), median absolute relative error (MARE), and median relative error (MRE) for single stock (SS) and multi-stock models (MS).}\label{fig:PMs}
\end{figure}

\end{landscape}

\newpage

\begin{landscape}

\begin{figure}[p]

{\centering \includegraphics[width=0.9\linewidth]{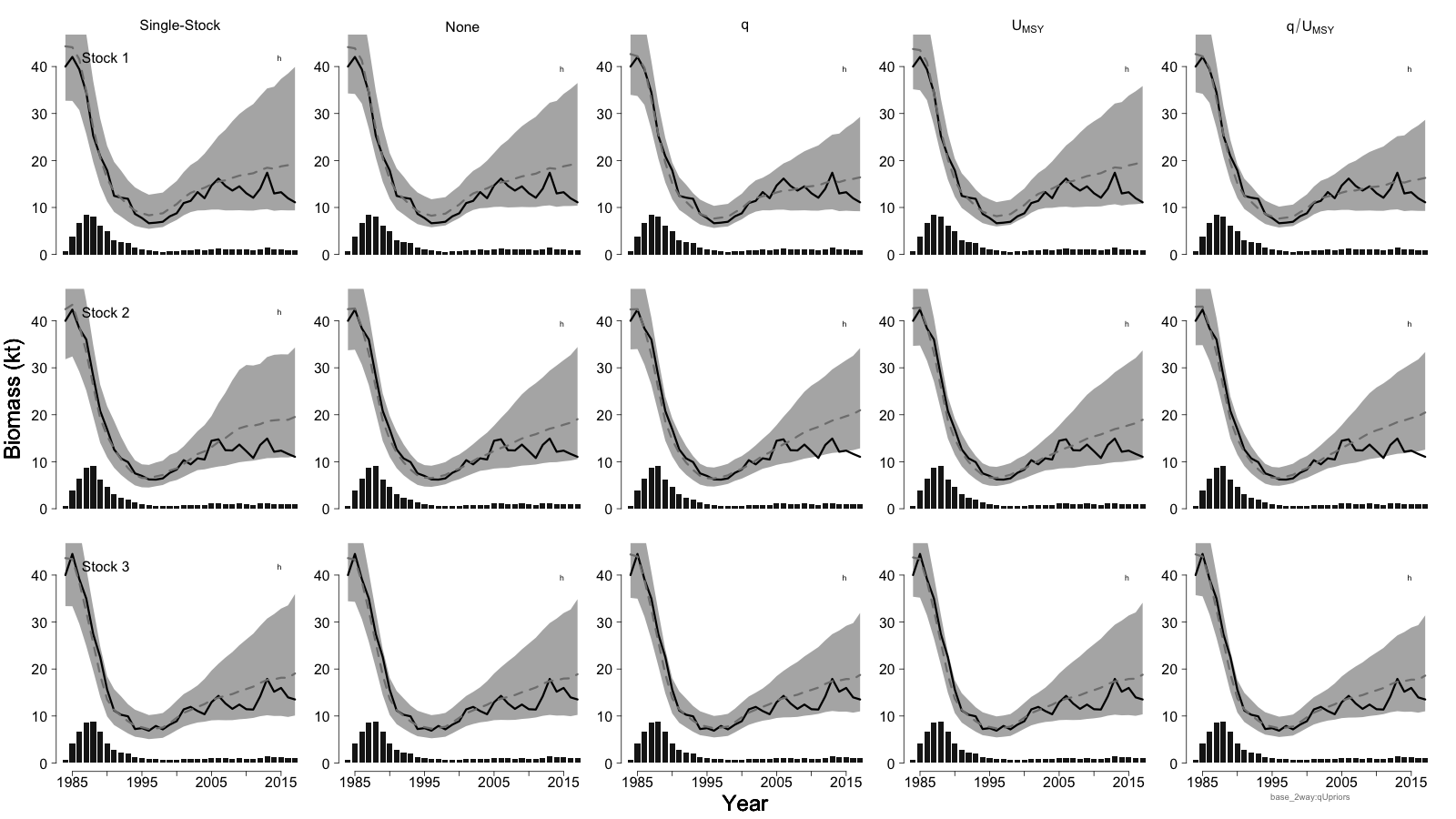} 

}

\caption{Time series of biomass and catch for a 3 stock complex, taken from a single simulation replicate using the base operating model. Thick unbroken lines indicate the simulated biomass values, while black vertical bars indicate the simulated catch. Assessment model estimated biomass is shown by dashed grey lines and 95\% confidence intervals by shaded regions. Single-stock estimates are in the first column and the remaining columns show the four multi-stock model configurations, with titles corresponding to which shared priors are estimated. The 95\% confidence intervals are calculated from the Hessian matrix for leading model parameters using the $\Delta$-method by TMB's ADREPORT() function.}\label{fig:BCsimBase}
\end{figure}

\end{landscape}

\newpage

\begin{landscape}

\begin{figure}[p]

{\centering \includegraphics[width=0.9\linewidth]{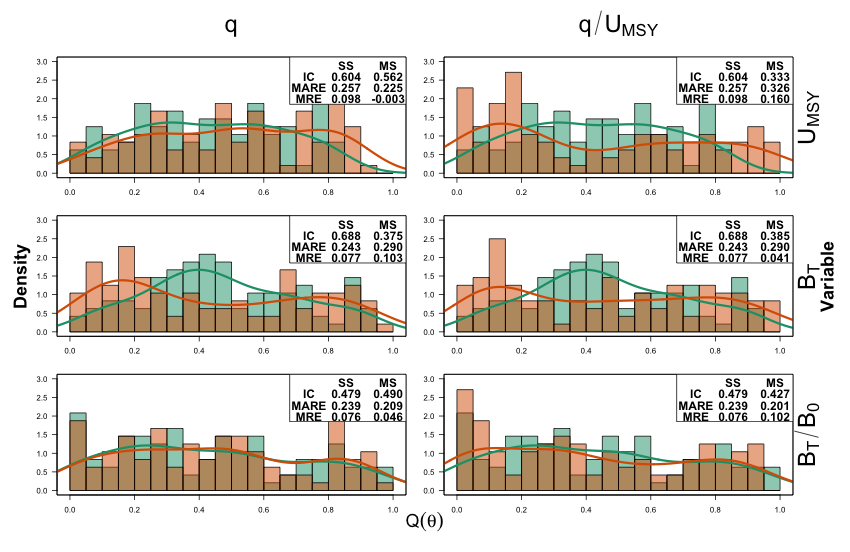} 

}

\caption{Density of predictive quantiles $Q(\theta)$ for estimates of key management parameters (rows) from single stock and $q$ and $q/U_{MSY}$ hierarchical multi-stock model configuration, fit to 4 identical stock under a 1-way trip fishing history over a long time-series of observations, initialised at unfished ($L = 0$). Bars show probability density of $Q$ distributions, with lines showing the kernel smoothed density for easier comparison between single stock (green) and multi-stock (orange) models. Top right hand corners of each panel show interval coverage (IC), median absolute relative error (MARE), and median relative error (MRE) for single stock (SS) and multi-stock models (MS).}\label{fig:PMOneWay}
\end{figure}

\end{landscape}

\newpage

\begin{landscape}

\begin{figure}[p]

{\centering \includegraphics[width=0.9\linewidth]{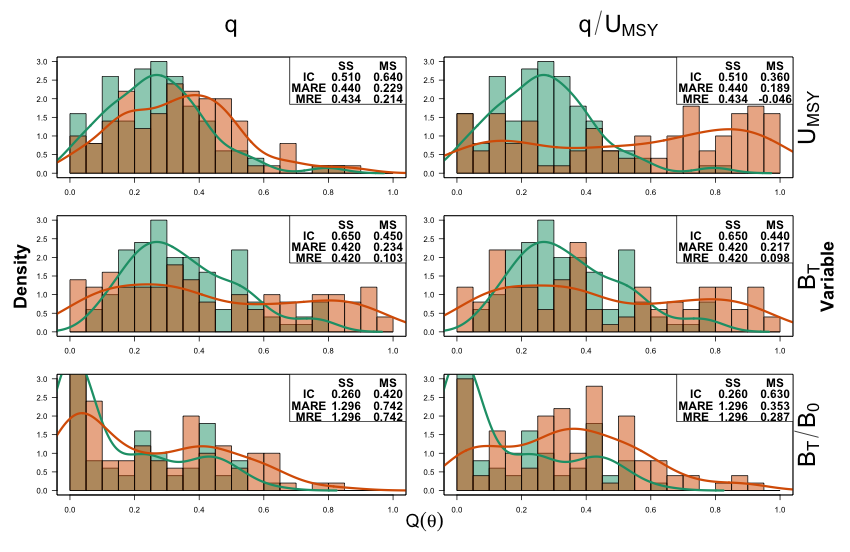} 

}

\caption{Density of predictive quantiles $Q(\theta)$ for estimates of key management parameters (rows) from single stock and $q$ and $q/U_{MSY}$ hierarchical multi-stock model configurations fit to a complex of four stocks with a 2-way trip fishing history with one low data quality stock ($L = 1$), which had a short time series of observations and was initialised at 40\% of unfished. Bars show probability density of $Q$ distributions, with lines showing the kernel smoothed density for easier comparison between single stock (green) and multi-stock (orange) models. Top right hand corners of each panel show interval coverage (IC), median absolute relative error (MARE), and median relative error (MRE) for single stock (SS) and multi-stock models (MS).}\label{fig:PMTwoWay}
\end{figure}

\end{landscape}

\newpage

\begin{landscape}

\begin{figure}[p]

{\centering \includegraphics[width=0.9\linewidth]{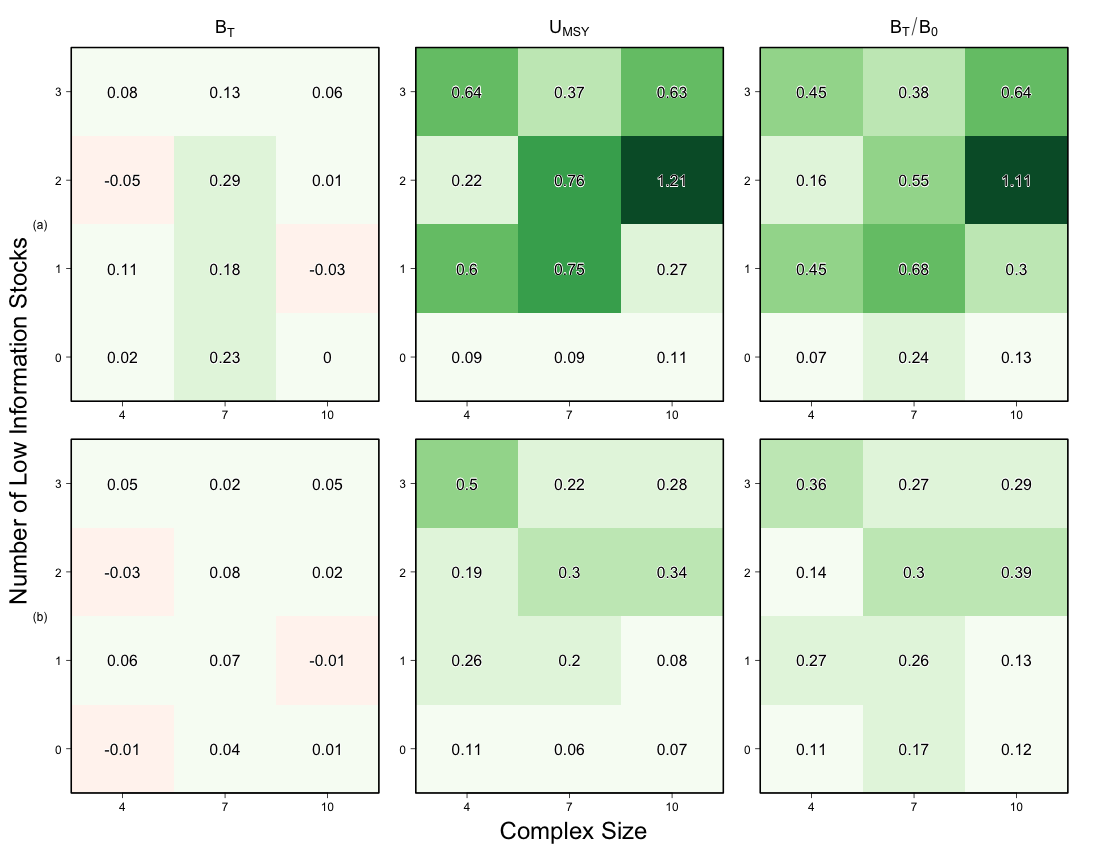} 

}

\caption{Response surface plots of (a) $\Delta(\theta_s) = \frac{MARE_{MS}(\theta)}{MARE_{SS}(\theta)} - 1$ and (b) $\overline\Delta(\theta) = \frac{\sum_s MARE_{MS}(\theta_s)}{\sum_s MARE_{SS}( \theta_s)} - 1$ values for $B_{1,T}$ (col. 1) and $U_{1,MSY}$ (col. 2) and $B_{1,T}/B_{1,0}$ (col. 3). Surfaces are plotted as responses to complex size $S$ along the horizontal axis, and number of low information stocks $L$ along the vertical axis. Colours represent the magnitude of the response value, with higher absolute values showing more saturation than absolute values closer to 0, and hue changing from red to green as responses pass from negative, indicating that the single stock performs better, to positive, indicating that the multi-stock model performs better. Response values in each cell are the mean response values for all experimental treatments where $S$ and $L$ took the corresponding values along the axes.}\label{fig:respSurface}
\end{figure}

\end{landscape}

\newpage

\begin{landscape}

\begin{figure}[p]

{\centering \includegraphics[width=0.9\linewidth]{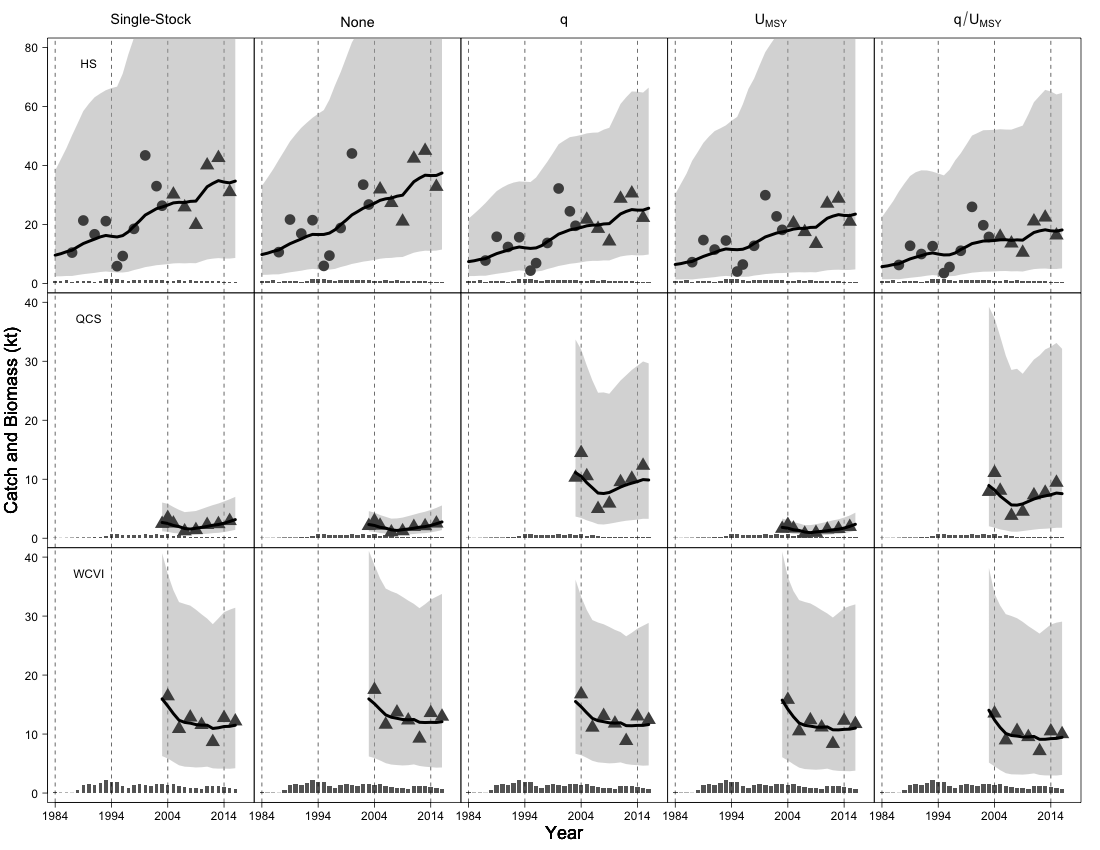} 

}

\caption{Estimated biomass time series for all three Dover Sole stocks. Estimates were produced by the single-stock and 4 top scoring multi-stock assessment model configurations under the low process error variance hypothesis. Grey regions indicate 95\% confidence intervals around the maximum likelihood estimates, indicated by the black lines. Black vertical bars at the bottom of each plot show absolute landings and discards. Points indicate survey biomass data scaled by estimated catchability. Circular data points indicate Survey 1 (HS only), while triangular points indicate Survey 2.}\label{fig:doverAssessPlot}
\end{figure}

\end{landscape}

\newpage

{\Large \textbf{Supplementary Material}}

\beginsupplement

\hypertarget{experimental-design}{%
\section{Experimental Design}\label{experimental-design}}

\hypertarget{latin-hyper-rectangle-designs}{%
\subsection{Latin Hyper-rectangle
Designs}\label{latin-hyper-rectangle-designs}}

To sample our experimental design space we used a space filling design
that we call the latin hyper-rectangle design (LHrD). The LHrD is a
modification of the latin hypercube design (LHD), which is a
multi-dimensional stratified random sampling approach (Kleijnen 2008).
There are a few differences between a classical LHD and our LHrD, but
the principle remains the same: LHrDs are a space filling design that
test every factor level while avoiding a full factorial design, reducing
computational overhead. This reduction in overhead is useful given that
multi-stock models are subject to the curse of dimensionality.

The main difference between a LHrD and a LHD is in how the design space
is sampled. In LHDs, the dimensions of the design space (the simulation
model inputs, or experimental factors) are split into an equal number of
segments - thus, a hypercube - and factor levels are sampled from within
each segment so that the resulting design has the latin property. In a
LHrD, the constraint that each dimension has an equal number of segments
is relaxed, forming a hyper-rectangle instead. Again, segments of each
dimension are sampled so that the resulting design has the latin
property.

There are some features of latin hyper-cubes that are lost when relaxing
the equal segments constraint to create latin hyper-rectangles. For
example, the dimensions that are broken into more segments have fewer
samples, due to the lower number of segments on the ``shorter''
dimensions of the hyper-rectangle. This is not a huge problem, however,
since given sufficiently many dimensions those segments will be sampled
in another slice of the hyper-rectangle. In this way, every factor level
is tested in combination with multiple levels of other factors, creating
a space filling design. We also restricted our design to a so-called
\emph{deterministic} LHrD, where the sampled elements within each
segment were the same for every treatment, instead of randomly sampled
from within that segment.

We found that defining our own approach that relaxed the hyper-cube
constraint was preferable for two main reasons. First, it's not always
necessary to test an equal number of levels for every factor. For
example, understanding the effects of one-way and two-way trips probably
only requires 2 levels of fishing intensity, whereas using a hypercube
design would have required us to arbitrarily define multiple extra
scenarios in between the extremes we've given in this paper. Second, we
considered the time spent on creating this process to be an investment
in a method that could be used in future simulation experiments as part
of our own work, and potentially the work of others.

Our process for defining a LHrD starts by systematically defining a
latin hyper-rectangle with the same number of dimensions as we have
factors. Each dimension of the hyper-rectangle is has a number of
entries equal to the number of levels for the corresponding factor, and
in this way the hyper-rectangle spans the experimental tableau. Sampling
the experimental tableau then reduces to sampling the entries of the
hyper-rectangle. We followed the steps below to create our design in the
R statistical programming language (available at
\url{https://github.com/samueldnj/LHrD}):

\begin{enumerate}
\def\labelenumi{\arabic{enumi}.}
\tightlist
\item
  Define experimental factors and choose factor levels. The number of
  factors \(F\) is the number of dimensions of the hyper-rectangle.
\item
  Choose the factor with the highest number \(j\) of levels. This is
  number of distinct values of hyper-rectangle entries.
\item
  Create an \(F\)-dimensional array \(A\), and assign each dimension of
  \(A\) to correspond to a different experimental factor by setting
  array dimensions to the number of levels of that factor.
\item
  Populate the entries of \(A\) with the integers \(\{0,1,2,...,j-1\}\)
  by adding the dimension indices modulo \(j\). For example, if
  \(F = 3\) and \(j = 4\), the entry in the \((1,1,1)\) position of
  \(A\) will be \(3 \mod 4 = 3\), and the entry in the \((1,2,1)\)
  position of \(A\) will be \(4 \mod 4 = 0\), etc.
\item
  In sequence, permute \(F-1\) dimensional slices of \(A\) by applying a
  random permutation to each dimension's indices. In R code, for the
  first dimension:
  \texttt{A{[}1:d\_1,,,,{]}\ \textless{}-\ A{[}sample(d\_1,d\_1),,,,{]}.}
\item
  To choose a sample of experimental treatments, randomly sample
  \(e \in \{0,1,..,j-1\}\). The array indices of each instance of \(e\)
  in \(A\) then correspond to levels of each factor, creating an
  experimental treatment.
\end{enumerate}

After step 4, the array \(A\) is a hyper-rectangular array with entries
\(0,1,...,j-1\) with the latin property. We do not prove this property
here, but the key is in the choice of the maximum dimension \(j\) of the
array to be the modulus of entries. Experimental designs can be formed
directly from this point using step 6, however in the experimental
design literature these designs are considered to be sub-optimal
{[}Kleijnen (2008); Ch 5{]}, as they favour diagonals of the
experimental tableau and may then cause dependence between treatments.
This is why we apply step 5, where we reduce dependence by randomising
entries with permutations, which preserve the latin property. In order
to maximise the randomisation of entries, we suggest using randomly
selected derangements (permutations that fix no points), however this
would require some extra machinery to produce (Martı'nez et al. 2008).

We relied heavily on the established properties of LHDs when developing
the LHrD, and as such have not conducted extensive tests on the
robustness of this approach. For example, classic LHDs have been refined
into ``maximin'' and ``nearly-orthogonal'' LHDs, which reduce some of
the interdependence between factor levels; however, we did not do this
here and leave it for future work or other analysts. Moreover, we did
not conduct meta-model validation in our experiments, except for an
informal ad-hoc validation when experiments were run multiple times.
Each new run produced new random samples of treatments, and the
resulting meta-models were often compared. A formal approach to
meta-model validation could be easily facilitated using the LHrD
framework: simply choose two entries, \(e_1, e_2\), and run both
designs. A meta-model fit to the results of the design based on \(e_1\)
could then be validated on the results of design based on \(e_2\), or
vice versa.

\hypertarget{our-experimental-design}{%
\subsection{Our Experimental Design}\label{our-experimental-design}}

We give the table for our experimental design below. We added
convergence metrics to the table for each scenario and AM configuration,
which describe how many attempts were made for each replicate, and how
many replicates in total were required to reach 100 converged replicates
in each combination.

\rowcolors{2}{gray!6}{white}
\begin{table}[!h]

\caption{\label{tab:makeExpDesignConvTable}The space filling experimental design used for the simulation experiments (columns 1-5), and the total number of simulation replicates required to get a full set of data for each hierarchical multi-stock assessment model configuration (columns 6-9).}
\centering
\resizebox{\linewidth}{!}{
\begin{tabular}[t]{ccccccccc}
\hiderowcolors
\toprule
\multicolumn{5}{c}{\bfseries Experimental Factor Levels} & \multicolumn{4}{c}{\bfseries Prior Configuration} \\
\cmidrule(l{2pt}r{2pt}){1-5} \cmidrule(l{2pt}r{2pt}){6-9}
Uhist & initYear & nS & initDep & nDiff & noJointPriors & qPriorOnly & UmsyPriorOnly & qUpriors\\
\midrule
\showrowcolors
c(0.2,4,1) & 1984 & 4 & 0.4 & 0 & 100 & 100 & 100 & 100\\
c(1,1,1) & 2003 & 7 & 0.4 & 0 & 102 & 102 & 102 & 102\\
c(1,1,1) & 1984 & 10 & 0.4 & 0 & 102 & 101 & 101 & 101\\
c(0.2,4,1) & 2003 & 10 & 0.4 & 0 & 100 & 100 & 100 & 100\\
c(1,1,1) & 1984 & 4 & 0.7 & 0 & 104 & 103 & 103 & 102\\
\addlinespace
c(0.2,4,1) & 2003 & 4 & 0.7 & 0 & 100 & 100 & 100 & 100\\
c(1,1,1) & 2003 & 10 & 0.7 & 0 & 102 & 101 & 101 & 101\\
c(1,1,1) & 2003 & 4 & 1.0 & 0 & 104 & 103 & 103 & 103\\
c(0.2,4,1) & 1984 & 7 & 1.0 & 0 & 100 & 100 & 100 & 100\\
c(1,1,1) & 2003 & 4 & 0.4 & 1 & 104 & 103 & 102 & 102\\
\addlinespace
c(0.2,4,1) & 1984 & 7 & 0.4 & 1 & 100 & 100 & 100 & 100\\
c(1,1,1) & 1984 & 7 & 0.7 & 1 & 104 & 102 & 101 & 102\\
c(0.2,4,1) & 2003 & 7 & 0.7 & 1 & 100 & 100 & 100 & 100\\
c(0.2,4,1) & 1984 & 10 & 0.7 & 1 & 100 & 100 & 100 & 100\\
c(0.2,4,1) & 1984 & 4 & 1.0 & 1 & 100 & 100 & 100 & 100\\
\addlinespace
c(1,1,1) & 2003 & 7 & 1.0 & 1 & 104 & 102 & 102 & 102\\
c(1,1,1) & 1984 & 10 & 1.0 & 1 & 103 & 101 & 101 & 101\\
c(0.2,4,1) & 2003 & 10 & 1.0 & 1 & 100 & 100 & 100 & 100\\
c(1,1,1) & 1984 & 4 & 0.4 & 2 & 102 & 101 & 102 & 102\\
c(0.2,4,1) & 2003 & 4 & 0.4 & 2 & 101 & 100 & 100 & 100\\
\addlinespace
c(1,1,1) & 2003 & 10 & 0.4 & 2 & 103 & 101 & 101 & 101\\
c(1,1,1) & 2003 & 4 & 0.7 & 2 & 102 & 102 & 101 & 102\\
c(0.2,4,1) & 1984 & 7 & 0.7 & 2 & 100 & 100 & 100 & 100\\
c(1,1,1) & 1984 & 7 & 1.0 & 2 & 102 & 102 & 102 & 101\\
c(0.2,4,1) & 2003 & 7 & 1.0 & 2 & 100 & 101 & 100 & 100\\
\addlinespace
c(0.2,4,1) & 1984 & 10 & 1.0 & 2 & 100 & 100 & 100 & 100\\
c(1,1,1) & 1984 & 7 & 0.4 & 3 & 102 & 101 & 101 & 101\\
c(0.2,4,1) & 2003 & 7 & 0.4 & 3 & 100 & 101 & 100 & 100\\
c(0.2,4,1) & 1984 & 10 & 0.4 & 3 & 100 & 100 & 100 & 100\\
c(0.2,4,1) & 1984 & 4 & 0.7 & 3 & 100 & 100 & 100 & 100\\
\addlinespace
c(1,1,1) & 2003 & 7 & 0.7 & 3 & 102 & 101 & 101 & 101\\
c(1,1,1) & 1984 & 10 & 0.7 & 3 & 103 & 101 & 101 & 101\\
c(0.2,4,1) & 2003 & 10 & 0.7 & 3 & 100 & 100 & 100 & 100\\
c(1,1,1) & 1984 & 4 & 1.0 & 3 & 104 & 103 & 102 & 102\\
c(0.2,4,1) & 2003 & 4 & 1.0 & 3 & 100 & 100 & 100 & 100\\
c(1,1,1) & 2003 & 10 & 1.0 & 3 & 102 & 101 & 101 & 101\\
\bottomrule
\end{tabular}}
\end{table}
\rowcolors{2}{white}{white}

\newpage

\hypertarget{meta-models-for-performance-metrics}{%
\section{Meta-models for performance
metrics}\label{meta-models-for-performance-metrics}}

Performance metrics were modeled as responses to experimental factors
and assessment model prior configurations using generalised linear
``meta-models''. Meta-modeling is a part of a formal approach to
simulation experimentation, where outputs of a complex simulation model
are viewed as responses to simpler functions of simulation model inputs
(Kleijnen 2008). The parameters of the simpler function, or meta-model,
are then used to improve interpretation of the results of complex
simulation experiments. We used generalised linear meta-models as they
are robust to heterogeneous variance of response variable residuals
(McCullagh 1984), which were common in our experimental treatment
outputs.

For performance metrics
\(y \in \{\Delta(\theta_s),\overline\Delta(\theta), MARE_{ss}(\theta_s), MARE_{ms}(\theta_s)\}\)
we estimated the coefficients \(\beta\) of a generalised linear model
\[ y = \beta_0 + \beta_{config} + \sum_{i} \beta_i x_i \] for each
experiment, where the \(\beta_0\) is the intercept, \(\beta_{config}\)
is the effect of the multi-stock assessment model prior configuration,
and the coefficients \(\beta_i\) are the factor effects for factor
levels \(x_i\). Numerical explanatory variables, such as the year of
initialisation and initial depletion, were scaled to \([-1,1]\) to allow
direct comparison of numeric effects with qualitative factor effects. To
reduce the number of experimental treatments, we sampled factor levels
using a space filling design (Kleijnen 2008). To reduce qualitative
factors, we fit the historical fishing intensity and initial year of
assessment as continuous variables even though they may not have
continuous, or even approximately linear responses. Our reasoning for
this is that both factors have 2 levels, and so a linear effect will
capture the difference between the level effects sufficiently. For the
historical fishing intensity, we regressed on the highest multiple of
\(U_{s,MSY}\) in the history, that is, \(U_d = 1\) for one-way trips,
and \(U_d = 2\) for two way trips.

In our experiment, the intercept term \(\beta_0\) represents the average
response value at the reference levels of qualitative factors in the
model. The only qualitative factor we use is the choice of multi-stock
model configuration, so the intercept of the \(\Delta(\theta_s)\) models
is \(\beta_0 = \beta_{None}\), representing the ``null'' multi-stock
assessment configuration. When there are no qualitative factors in the
meta-model, such as in the \(MARE_{ss}(\theta_1)\) models, \(\beta_0\)
is simply the average response value over all factors.

\newpage

\hypertarget{assessment-model-structure}{%
\section{Assessment Model Structure}\label{assessment-model-structure}}

In what follows, we denote by \(\hat{x}\) the estimate of a derived or
leading model parameter \(x\).

\hypertarget{biomass-dynamics}{%
\subsection{Biomass dynamics}\label{biomass-dynamics}}

We minimized the effect of assessment model mis-specification by
matching the deterministic components of the biomass dynamics in the
assessment models and the operating model (Eq. 1). In all assessment
model configurations the leading biological parameters were
\(B_{s,MSY} = B_{s,0}/2\) and \(U_{s,MSY} = r_s/2\). Biomass time series
in the assessment models were initialized at time \(T_1\), either at
unfished levels \(B_{s,T_1} = B_{s,0}\), or at a separately estimated
non-equilibrium value \(\hat{B}_{s,T_1}\) when \(T_1 > 1984\) or the
initial simulated biomass was below unfished levels. Biomass parameters
were penalized by normal prior distributions centered at or near their
corresponding true values, i.e., \[
\begin{split}
\hat{B}_{s,MSY} \sim N(B_{s,MSY}, B_{s,MSY}), \\
\hat{B}_{s,T_1} \sim  N(B_{s,MSY}/2, B_{s,MSY}/2),
\end{split}
\] which allows estimates to vary within a realistic range by giving
each prior a relative standard deviation of 100\%.

\hypertarget{productivity-prior}{%
\subsubsection{Productivity prior}\label{productivity-prior}}

When we jointly modeled stock-specific optimal harvest rates
\(U_{s,MSY}\), we assumed \(U_{s,MSY}\) values shared a log-normal
distribution with estimated hyperparameters (U.1, Table 3). The
estimated prior mean \(\hat{\overline{U}_{MSY}}\) followed a normal
hyperprior (U.2, Table 3) where \(m_U\) was randomly drawn from a log
normal distribution with a mean of \(0.08\) and a standard deviation
corresponding to a 20\% coefficient of variation. The hyperprior
standard deviation \(s_U = 0.08\) was chosen to give a roughly 100\% CV
in the hyperprior for the prior mean and allow the stock-specific values
affect the estimate of \(\hat{\overline{U}_{MSY}}\) more than the
hyperprior. We chose a normal prior as this is the least informative
while remaining continuous across the whole domain of the parameter
space.

The estimated prior variance followed an inverse gamma distribution
(U.3, Table 3) with \(\alpha_U = 1\) and \(\beta_U = 0.34\), to induce a
log-normal coefficient of variation of 20\% in the shared \(U_{MSY}\)
prior. We chose this prior structure recognizing the shared biology of
dover sole stocks implies productivities of similar magnitude (Myers et
al. 1999). When we modeled the optimal harvest rates separately we
assumed optimal harvest rate parameters followed the hyperprior directly
(U.4, Table 3) with the same \(m_U\) and \(s_U\) values.

\hypertarget{observation-models}{%
\subsection{Observation models}\label{observation-models}}

Similar to the biomass dynamics, we matched the observational model
structures for the operating and assessment models (Equation 3). A
shared species-level prior distribution was defined for stock-specific
catchabilities \(q_{o,s}\), with between stock variance
\(\iota^2_{q,o}\). Informative priors were also defined for survey
observation error variances \(\tau_o^2\), with hyperparameters chosen so
that the prior modes were equal to the simulated values for each survey.

\hypertarget{catchability}{%
\subsubsection{Catchability}\label{catchability}}

When we estimated the hierarchical prior on catchability parameters, we
used the same model structure for the prior as the simulated
catchability model for each survey. Estimates of stock-specific
catchability \(\hat{q}_{o,s}\) were drawn from a shared log-normal
distribution with estimated hyperparameters \(\hat{\bar{q}}_o\) and
\(\hat\iota_{q,o}^2\) (q.1, Table 3). The estimated prior mean followed
a normal hyperprior where \(m_q\) was drawn randomly as with the
\(U_{MSY}\) prior, with an average of \(0.55\) (the average of the two
surveys) and 20\% CV, while \(v_q = 0.55\) for a 100\% hyperprior CV
(q.2, Table 3), for similar reasoning as the \(U_{MSY}\) priors.

The prior variance followed an inverse gamma distribution (q.3, Table
3). Inverse gamma hyperparameters \(\alpha_q = 1\) and
\(\beta_q = 0.34\) were chosen induce a \(\hat{\iota}_{q,o}^2\) value
that corresponds to a 20\% CV, inducing a shrinkage effect. This prior
structure reflects an assumption that dover Sole stocks have a similar
availability to survey gear based on similar habitat preferences. Like
the productivity prior, when catchability was estimated without a shared
prior, we bypassed the mid-level prior, penalizing stock-specific
catchability using the normal hyperprior with the same \(m_q\) and
\(v_q\) values to signify no change in prior information (q.4, Table 3).
We chose a normal hyperprior because it is less informative than a
log-normal distribution, and the mean \(m_q = .6\) and variance
\(v_q = 0.36\) are chosen to produce a relative standard deviation of
100\%, allowing \(\bar{q}_o\) and \(q_{o,s}\) to vary in a realistic
range, but informative enough to induce a shrinkage effect.

\hypertarget{observation-errors}{%
\subsubsection{Observation errors}\label{observation-errors}}

Observation errors for each survey were assumed to be drawn from a
single log-normal distribution across stocks, with estimated
log-standard deviation \(\hat{\tau}_o\). To improve convergence in
repeated simulation trials, we assumed the estimated log-variance
\(\hat{\tau}_o^2\) followed an inverse gamma prior distribution
\begin{equation}
\hat{\tau}_o^2 \sim IG(\alpha_{\tau_o},\beta_{\tau_o}).
\end{equation} \noindent Like the process error variance priors, the
hyperparameter \(\beta_{\tau_o}\) was chosen to place the mode of the
inverse gamma distribution at the simulated values of \(\tau_o^2\) when
\(\alpha_{\tau_o} = 0.1\).

\newpage

\hypertarget{extra-performance-metrics-for-the-base-operating-model}{%
\section{Extra performance metrics for the base operating
model}\label{extra-performance-metrics-for-the-base-operating-model}}

\begin{figure}[p]

{\centering \includegraphics[width=0.6\linewidth]{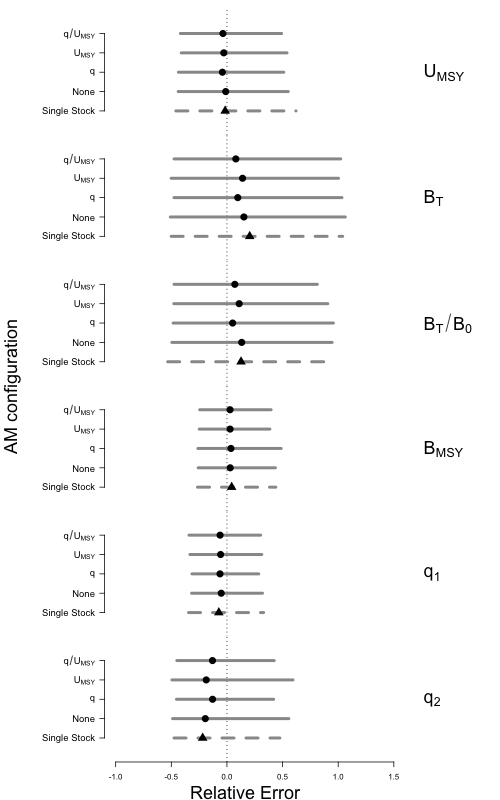} 

}

\caption{Relative error distributions for stock 1 leading and derived parameters estimated by the single stock (dashed lines and triangular points) and 4 multi-stock assessment models (solid lines and circular points) fit to data from the base operating model. Points indicate median relative errors and the grey lines the central 95\% of the relative error distribution. From the top, parameters are optimal exploitation rate ($U_{MSY}$), terminal biomass ($B_{T}$), optimal equilibrium biomass ($B_{MSY}$), terminal biomass relative to unfished ($B_{T}/B_{0}$), and catchability from surveys 1 ($q_{1}$) and 2 ($q_{2}$). Assessment model (AM) configurations indicate the single stock model, or the parameters that had hierarchical prior distribution hyperparameters estimated in the multi-stock assessment model (e.g, $q/U_{MSY}$ indicates that shared priors on both catchability and productivity were estimated).}\label{fig:mreDists}
\end{figure}

\newpage

\begin{figure}[p]

{\centering \includegraphics[width=0.6\linewidth]{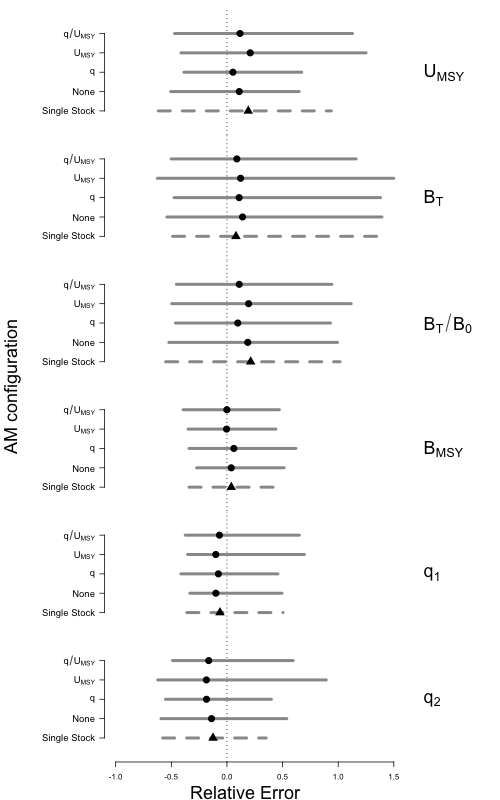} 

}

\caption{Relative error distributions for stock 1 leading and derived parameters estimated by the single stock (dashed lines and triangular points) and 4 multi-stock assessment models (solid lines and circular points) fit to data from the base operating model, but with a one-way trip fishing history. Points indicate median relative errors and the grey lines the central 95\% of the relative error distribution. From the top, parameters are optimal exploitation rate ($U_{MSY}$), terminal biomass ($B_{T}$), optimal equilibrium biomass ($B_{MSY}$), terminal biomass relative to unfished ($B_{T}/B_{0}$), and catchability from surveys 1 ($q_{1}$) and 2 ($q_{2}$). Assessment model (AM) configurations indicate the single stock model, or the parameters that had hierarchical prior distribution hyperparameters estimated in the multi-stock assessment model (e.g, $q/U_{MSY}$ indicates that shared priors on both catchability and productivity were estimated).}\label{fig:mreDistsOneWay}
\end{figure}

\newpage

\blandscape

\begin{figure}[p]

{\centering \includegraphics[width=0.9\linewidth]{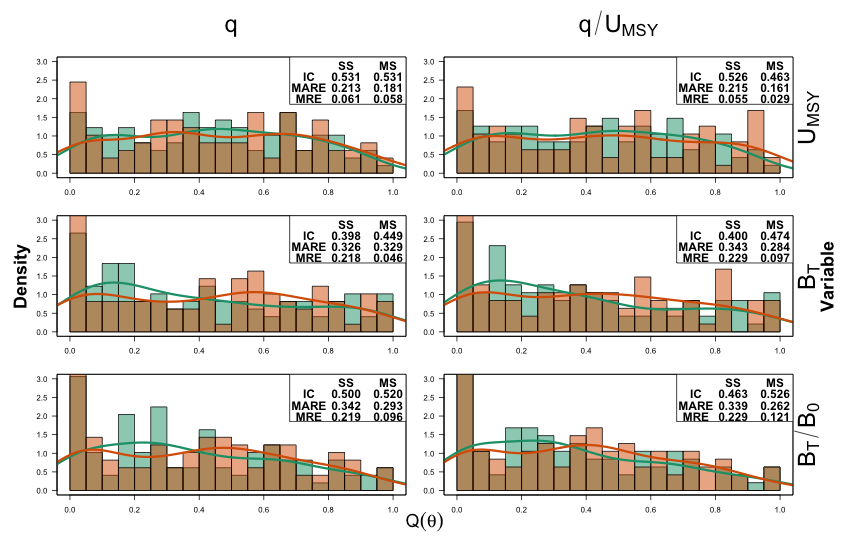} 

}

\caption{Density of predictive quantiles $Q(\theta)$ for estimates of key management parameters (rows) from single stock and $q$ and $q/U_{MSY}$ hierarchical multi-stock model configuration under the base operating model. Bars show probability density  of $Q$ distributions, with lines showing the kernel smoothed density for easier comparison between single stock (green) and multi-stock (orange) models. Top right hand corners of each panel show interval coverage (IC), median absolute relative error (MARE), and median relative error (MRE) for single stock (SS) and multi-stock models (MS).}\label{fig:PMsTwoWay}
\end{figure}

\elandscape

\newpage

\blandscape

\begin{figure}[p]

{\centering \includegraphics[width=0.9\linewidth]{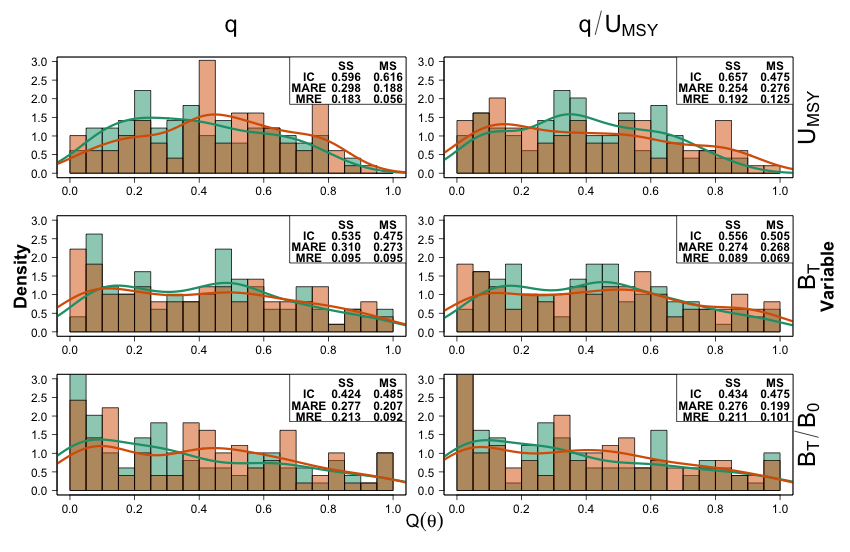} 

}

\caption{Density of predictive quantiles $Q(\theta)$ for estimates of key management parameters (rows) from single stock and $q$ and $q/U_{MSY}$ hierarchical multi-stock model configuration under the base operating model. Bars show probability density  of $Q$ distributions, with lines showing the kernel smoothed density for easier comparison between single stock (green) and multi-stock (orange) models. Top right hand corners of each panel show interval coverage (IC), median absolute relative error (MARE), and median relative error (MRE) for single stock (SS) and multi-stock models (MS).}\label{fig:PMsOneWay}
\end{figure}

\elandscape

\newpage

\hypertarget{assessments-of-british-columbias-dover-sole}{%
\section{Assessments of British Columbia's dover
sole}\label{assessments-of-british-columbias-dover-sole}}

\blandscape

\begin{figure}[p]

{\centering \includegraphics[width=0.9\linewidth]{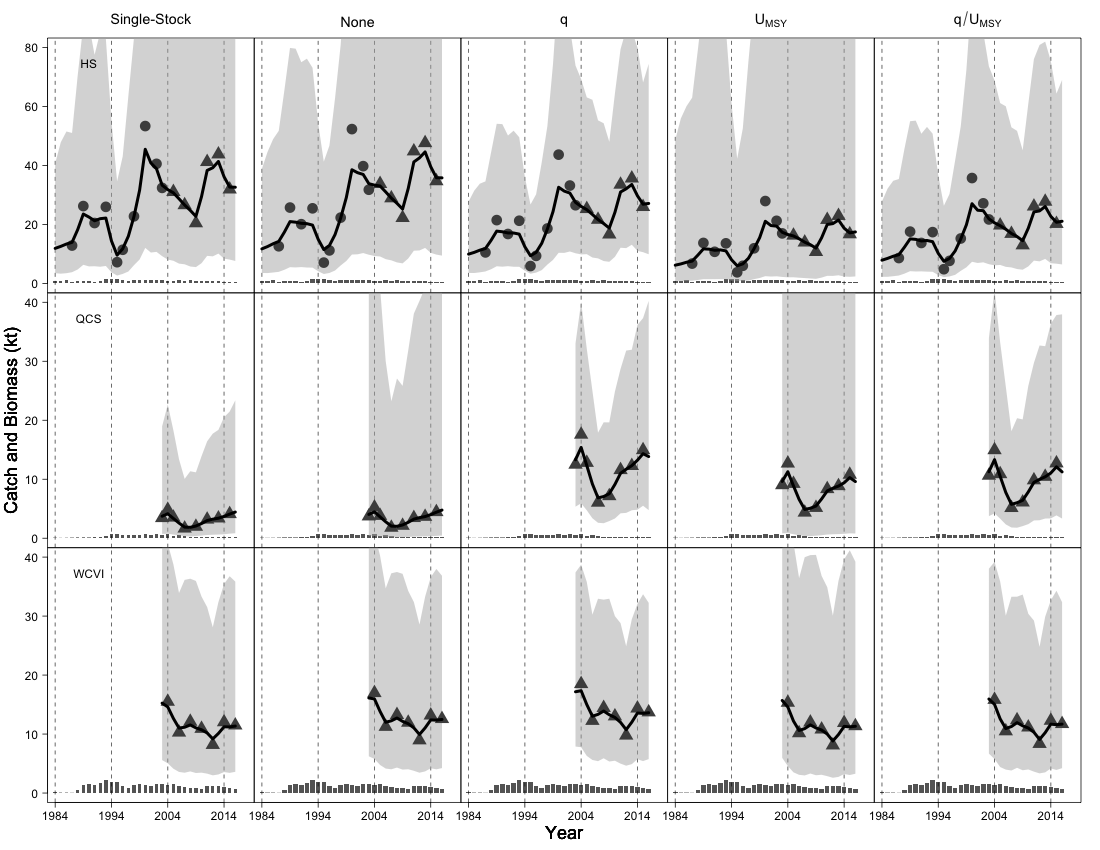} 

}

\caption{Estimated biomass time series for all three dover sole stocks. Estimates were produced by the single-stock and 4 top scoring multi-stock assessment model configurations under the high process error variance hypothesis. Grey regions indicate $95\%$ confidence intervals around the maximum likelihood estimates, indicated by the black lines. Grey bars at the bottom of each plot show absolute landings and discards. Points indicate survey biomass data scaled by estimated catchability. Circular data points indicate Survey 1 (HS only), while triangular points indicate Survey 2.}\label{fig:doverFitsHighPEPlot}
\end{figure}

\elandscape

\newpage

\hypertarget{references}{%
\section*{References}\label{references}}
\addcontentsline{toc}{section}{References}

\hypertarget{refs}{}
\leavevmode\hypertarget{ref-kleijnen2008design}{}%
Kleijnen, J.P. 2008. Design and analysis of simulation experiments.
Springer.

\leavevmode\hypertarget{ref-martinez2008generating}{}%
Martı'nez, C., Panholzer, A., and Prodinger, H. 2008. Generating random
derangements. \emph{In} 2008 proceedings of the fifth workshop on
analytic algorithmics and combinatorics (analco). SIAM. pp. 234--240.

\leavevmode\hypertarget{ref-mccullagh1984generalized}{}%
McCullagh, P. 1984. Generalized linear models. European Journal of
Operational Research \textbf{16}(3): 285--292. Elsevier.

\leavevmode\hypertarget{ref-myers1999maximum}{}%
Myers, R.A., Bowen, K.G., and Barrowman, N.J. 1999. Maximum reproductive
rate of fish at low population sizes. Canadian Journal of Fisheries and
Aquatic Sciences \textbf{56}(12): 2404--2419. NRC Research Press.

\end{document}